\documentclass[%
 reprint,
 amsmath,amssymb,
 aps,
]{revtex4-1}

\usepackage{graphicx}
\usepackage{dcolumn}
\usepackage{bm}
\usepackage{hyperref}
\begin{document}

\title{A solution to electric-field screening in diamond quantum electrometers}
\author{L. M. Oberg, M. O. de Vries, L. Hanlon, K. Strazdins, M. S. J. Barson}
\affiliation{Laser Physics Center, Research School of Physics, Australian National University, Australian Capital Territory 2601, Australia}
\author{J. Wrachtrup}
\affiliation{3rd Institute of Physics, University of Stuttgart and Institute for Quantum Science and Technology (IQST), Pfaffenwaldring 57, D-70569, Stuttgart, Germany}
\author{M. W. Doherty}
\email{marcus.doherty@anu.edu.au}
\affiliation{Laser Physics Center, Research School of Physics, Australian National University, Australian Capital Territory 2601, Australia}
\date{\today}
\begin{abstract}
There are diverse interdisciplinary applications for nanoscale resolution electrometry of elementary charges under ambient conditions. These include characterization of 2D electronics, charge transfer in biological systems, and measurement of fundamental physical phenomena. The nitrogen-vacancy center in diamond is uniquely capable of such measurements, however electrometry thus far has been limited to charges within the same diamond lattice. It has been hypothesized that the failure to detect charges external to diamond is due to quenching and surface screening, but no proof, model, or design to overcome this has yet been proposed. In this work we affirm this hypothesis through a comprehensive theoretical model of screening and quenching within a diamond electrometer and propose a solution using controlled nitrogen doping and a fluorine-terminated surface. We conclude that successful implementation requires further work to engineer diamond surfaces with lower surface defect concentrations.
\end{abstract}

\maketitle
\section*{Introduction}
Nanoscale charge imaging has been employed for diverse purposes including high-sensitivity biological and chemical sensors\cite{Cui2001,Patolsky2006}, detectors within quantum devices\cite{Eizerman2004}, and investigating fundamental physical phenomena\cite{Martin2008,Yoo1997}. Many techniques have been employed for precision electrometry with nanometer spatial resolution\cite{Yoo1997,Martin2008,Henning1995,Williams1989}, elementary charge detection\cite{Yoo1997,Martin2008,Devoret2000,Schonenberger1990,Martin1988,Cleland1998,Bunch2007,Salfi2010,Lee2008}, and the ability to operate at ambient temperatures and pressures\cite{Bunch2007,Lee2008}. However, no device yet possess all three of these properties simultaneously. This capability would be extremely valuable for investigating biological systems, such as neurons\cite{Barry2016,hanlon2019diamond}, and as a critical characterization tool for the emerging field of two-dimensional electronics\cite{Schwierz2010,Radisavljevic2011,Mak2016}. For example, atomic-resolution imaging of silicene to aid development of a room-temperature transistor\cite{Tao2015}, probing novel charged quasiparticles in MOS${}_2$ films\cite{Mak2012}, and detection of polarization skyrmions\cite{Das2019}.

The nitrogen-vacancy (NV) center\cite{Doherty2013} is currently the only system capable of nanoscale resolution electrometry of elementary charges under ambient conditions. This point defect in diamond consists of a substitutional nitrogen atom ($\text{N}_\text{S}$) situated adjacent to a carbon vacancy. Single NV centers have demonstrated room-temperature a.c. (d.c.) electric-field sensitivities reaching 202~$\text{V cm}^{-1}\text{ Hz}^{-1/2}$ (891~$\text{V cm}^{-1}\text{ Hz}^{-1/2}$)\cite{Dolde2011a,Dolde2014a}. Ensembles of NV centers have achieved shot-noise limited a.c. sensitivities on the order of 1~$\text{V cm}^{-1}\text{ Hz}^{-1/2}$\cite{Chen2017} and also been employed as in-situ electric field sensors within semiconductor heterojunctions\cite{Iwasaki2017}.

The NV center's proficiency for quantum sensing is due to a unique combination of capabilities. Firstly, the NV exhibits bright optical fluorescence allowing for identification of single defects that can be employed for measurements with nanoscale resolution. Secondly, the NV possesses a mechanism for optical spin initialization and readout which permits spin resonances of individual defects to be measured with high fidelity\cite{Doherty2013}. Finally, the NV boasts the longest room-temperature coherence time for any solid-state defect\cite{Balasubramanian2009} allowing for high-resolution detection of spin resonances when combined with optical readout. In addition to electrometry these properties have been applied for precision nano-magnetometry\cite{Tetienne2014a,Thiel2016,Haberle2015,Arai2015}, thermometry\cite{Kucsko2013,Neumann2013,Toyli2013} and quantum computing\cite{Waldherr2014b}, as well as proposed for investigating fundamental physical phenomena such as magnetic phase changes\cite{Cai2013} and coherent quantum transport\cite{ObergLachlanM2019}.

While the NV center can exist in several charge states, including neutral (NV${}^0$) and negative (NV${}^-$), only the latter possesses the aforementioned properties needed for electrometry. In particular the spin resonances of the NV${}^-$ triplet ground state are susceptible to electric field induced changes in its electron spin-spin interactions. The resulting Stark shifts can be detected using optically detected magnetic resonance (ODMR) and used to determine the field magnitude. Furthermore, vector components of the electric field can be measured through rotation of a bias transverse magnetic field\cite{Dolde2014a}.

Although single NV centers possess sufficiently high sensitivities, elementary charges external to diamond have not been detected with nanoscale resolution. We hypothesize that this is because of electric field screening and charge-state quenching of NV centers. For the former, recent experimental works have identified multiple screening sources inherent to diamond systems. These include Debye screening from bulk defects\cite{Broadway2018}, charge reorganization in primal sp${}^2$ surface defects\cite{Stacey2019} and polarization of adsorbed water vapor\cite{Mertens2016}. Moreover, p-type defects within bulk diamond\cite{Liao2008}, surface acceptor defects\cite{Stacey2019} and surface terminations with negative electron affinities\cite{Hauf2011} are known to quench the negative charge state, particularly for near-surface NV centers\cite{Cui2013,Ohno2012a}. However, the extent that these sources impact charge detection are unknown, and a comprehensive theoretical treatment is needed. 

The first aim of this paper is to develop a physical model of screening due to the external environment, internal diamond, and diamond surface. This is performed in Section~\ref{sec:general} where we identify that screening due to charge rearrangement amongst sp${}^2$ surface defects is the greatest impediment to electrometry. We propose a solution by saturating the charge traps through a sacrificial $\delta$-doped layer of $\text{N}_\text{S}$. The effectiveness of this idea is explored in Section~\ref{sec:cap} in which an analytical toy model is developed for a simplified electrometer. Finally, in Section~\ref{sec:design} this toy model is adapted into a more sophisticated device compatible with NV quantum sensing. The electrostatic properties of the device are modeled computationally, and the physical parameters optimized for charge detection. We conclude that this design successfully mitigates screening for concentrations of sp${}^2$ surface defects below approximately $10^{16}$~$\text{m}^{-2}$, two orders of magnitude lower than that currently demonstrated on fluorine-passivated diamond\cite{Stacey2019}.

\section{Screening and quenching within diamond}
\label{sec:general}

\begin{figure*}[]
	\centering
	\includegraphics[width=0.8\textwidth]{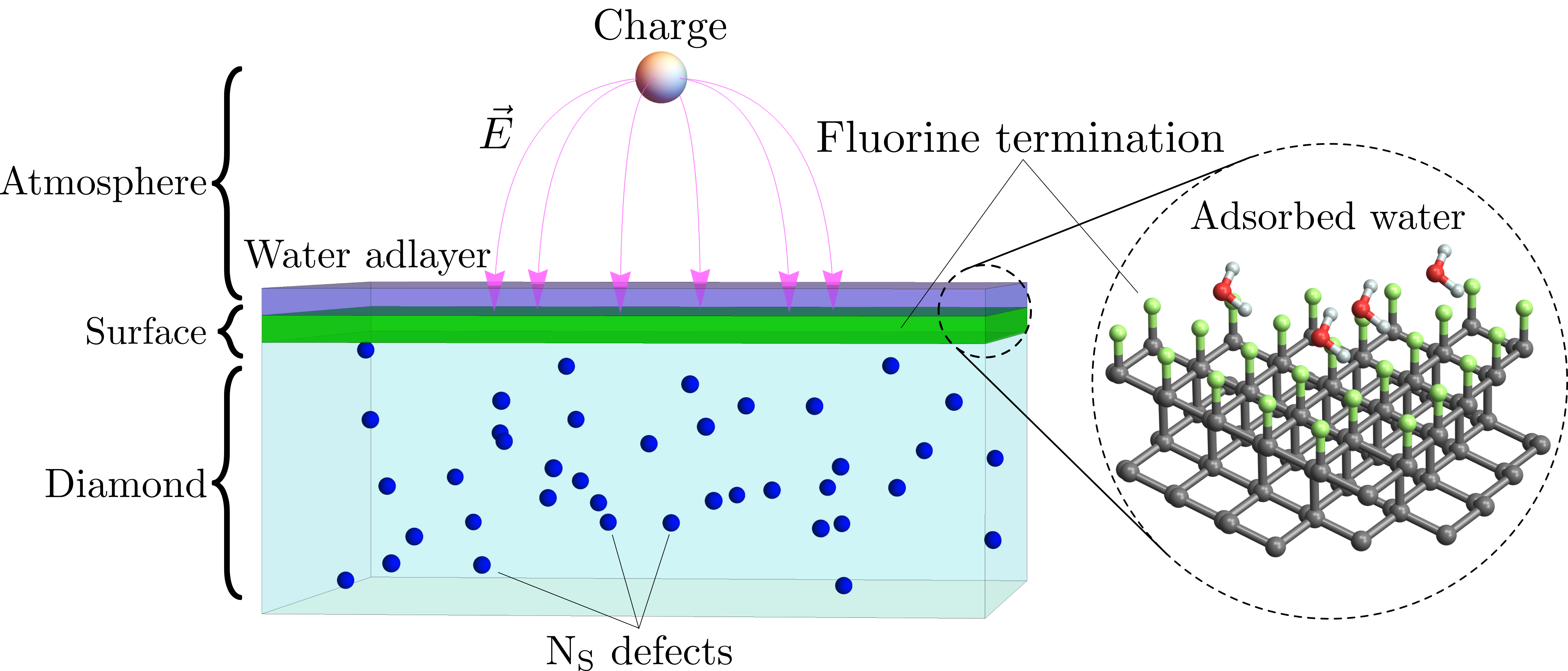}
	\caption{Schematic of the three-layered model used to investigate field screening and NV${}^-$ charge quenching. The external environment contains the ambient atmosphere and the charge distribution to be detected. Water vapor adsorbs to the diamond surface to form an adlayer which causes dielectric screening. This can be mitigated by passivating the diamond with fluorine which exhibits strong dipolar hydrophobicity\cite{Mayrhofer2016,Kissa2011}. Furthermore, fluorine-terminated diamond is chemically inert, room-temperature stable, and possesses a positive electron affinity\cite{Rietwyk2013}. The surface itself contains a high density of sp${}^2$ surface defects which introduce acceptor levels into the diamond band structure\cite{Stacey2019}. These readily quench the NV${}^-$ charge state and at partial occupation lead to strong surface screening effects. Within the diamond, uncontrolled doping of n-type defects such as $\text{N}_\text{S}$ lead to Debye screening\cite{FRS2013}.}
	\label{fig:regions}
\end{figure*}

Screening and quenching effects within diamond can be decomposed into three coupled systems; the external atmosphere, the surface and the internal diamond. These three environments and their associated screening/quenching sources are depicted in Figure~\ref{fig:regions}. In the following subsections we model each system individually and assess the associated impacts for diamond electrometry.

\subsection{External atmosphere}

We first address screening due to the external atmosphere at ambient temperature and pressure. As air possesses a relative permittivity of approximately unity it causes negligible electrical screening. In contrast, the physisorption of water vapor on the diamond surface may be a detrimental source of screening given water's high relative permittivity. This subsection reviews the current understanding of wetting on diamond surfaces and applies these results to form a cohesive theory of screening due to water adsorption.

Considering the most common diamond surface terminations, oxygen is known to be strongly hydrophilic and therefore not suitable for precision electrometry\cite{Widmann2014}. While hydrogen terminated surfaces exhibit polar hydrophobicity\cite{Mayrhofer2016,Mertens2016}, they are not viable for electrometry with near-surface NV centers. This is because they possess a negative electron affinity which introduces subsurface holes that quench the NV${}^-$ charge state\cite{Hauf2011}. However, this is not the case for fluorine-terminated diamond, which is chemically inert and room-temperature stable\cite{Cui2013} with a positive electron affinity\cite{Rietwyk2013}. Importantly, the fluorine surfaces exhibits strong hydrophobicity\cite{Widmann2014,Kissa2011}, reflected by its high wetting angle and a small physisorption energy of 0.07~eV for F-C(111) as determined using \textit{ab-initio} calculations\cite{Mayrhofer2016}. Ideally these adsorption energies could be used in conjunction with a suitable isotherm equation to determine the water coverage. Unfortunately, this is not possible as current studies neglect Gibbs contributions to adsorption and so cannot accurately predict coverages under ambient conditions.

Regardless, it can be expected that surface coverage is much lower than a monolayer. Simple analysis with the Brunauer-Emmett-Teller isotherm indicates negligible coverage as the condensation energy of water far exceeds that of physisorption (neglecting entropic and enthalpic contributions)\cite{Brunauer1938}. Comparisons can also be made to a recent spectroscopic study of water coverage on hydrophobic H-Si(111), which shares a similar wetting angle to F-C(111)\cite{Mertens2016,Widmann2014,Mayrhofer2016,Lange2009,Silvestrelli2006}. This observed that at ambient temperatures and pressures, surface coverage is an increasing function of humidity that never exceeds a monolayer\cite{Chen2018}. Microscopically, wetting behaviour on hydrophobic surfaces is far more complicated. Graphene templating in conjunction with atomic force microscopy has revealed complex water structures on H-Si(111), with nano-droplets up to 20~nm wide and 0.5~nm in height accumulating on surface defects and step edges\cite{Cao2011}.

We now investigate whether adsorbed water produces a major or minor screening effect. Given the ambiguity of the nature and extent of wetting on hydrophobic diamond surfaces, we will model the dielectric permittivity as a function of adsorbed water adlayers. This is denoted by $\theta$, the fractional amount of water monolayers adsorbed on the surface between 0 and 1. While surface-adsorbed vapor likely possesses a high rate of diffusion at room temperature, here we only consider the time-averaged response of the induced permittivity. As the electrical properties of the F-C(111) surface have been well-characterized theoretically\cite{Mayrhofer2016}, it shall be considered as the model surface for electrometry. However, the results presented here are likely applicable to other cuts of fluorine-terminated diamond surfaces as well. The polar C-F surface bonds induce an electric field which orientates the dipoles of the physisorbed water. This generates a net polarization in the direction of the C-F dipoles which can be calculated at thermal equilibrium. The polarization density is given by\cite{Atkins2014}
\begin{equation}\label{pol}
\vec{P}(\vec{E}_s)=P(\beta\mu |\vec{E}_s|)\hat{E}_s,
\end{equation}
where $\mu = 1.85$~D is the dipole moment of water, $\beta = (k_B T)^{-1}$, $\vec{E}_s$ is the electric field generated by the diamond surface and
\begin{equation}
P(x) = \mu \mathcal{N}\mathcal{L}(x),
\end{equation}
for $\mathcal{N}$ the number density of water molecules and $\mathcal{L}(x)$ the Langevin function given by
$$\mathcal{L}(x) = \coth(x) - \frac{1}{x}.$$

A value for $\mathcal{N}$ may be estimated by treating each F as a single adsorption site. This is justified as \textit{ab-initio} calculations indicate that the energy of a physisorbed $\text{H}_2\text{O}$ molecule is minimized when aligned laterally with a terminating F atom\cite{Mayrhofer2016}. In the direction perpendicular to the surface (denoted $\hat{z}$) we may bound the linear density of water molecules by that of liquid water, yielding $\mathcal{N} = 0.156\cdot\theta$ $\text{molecules/\AA}^3$. 

\begin{figure}[]
	\centering
	\includegraphics[width=0.5\textwidth]{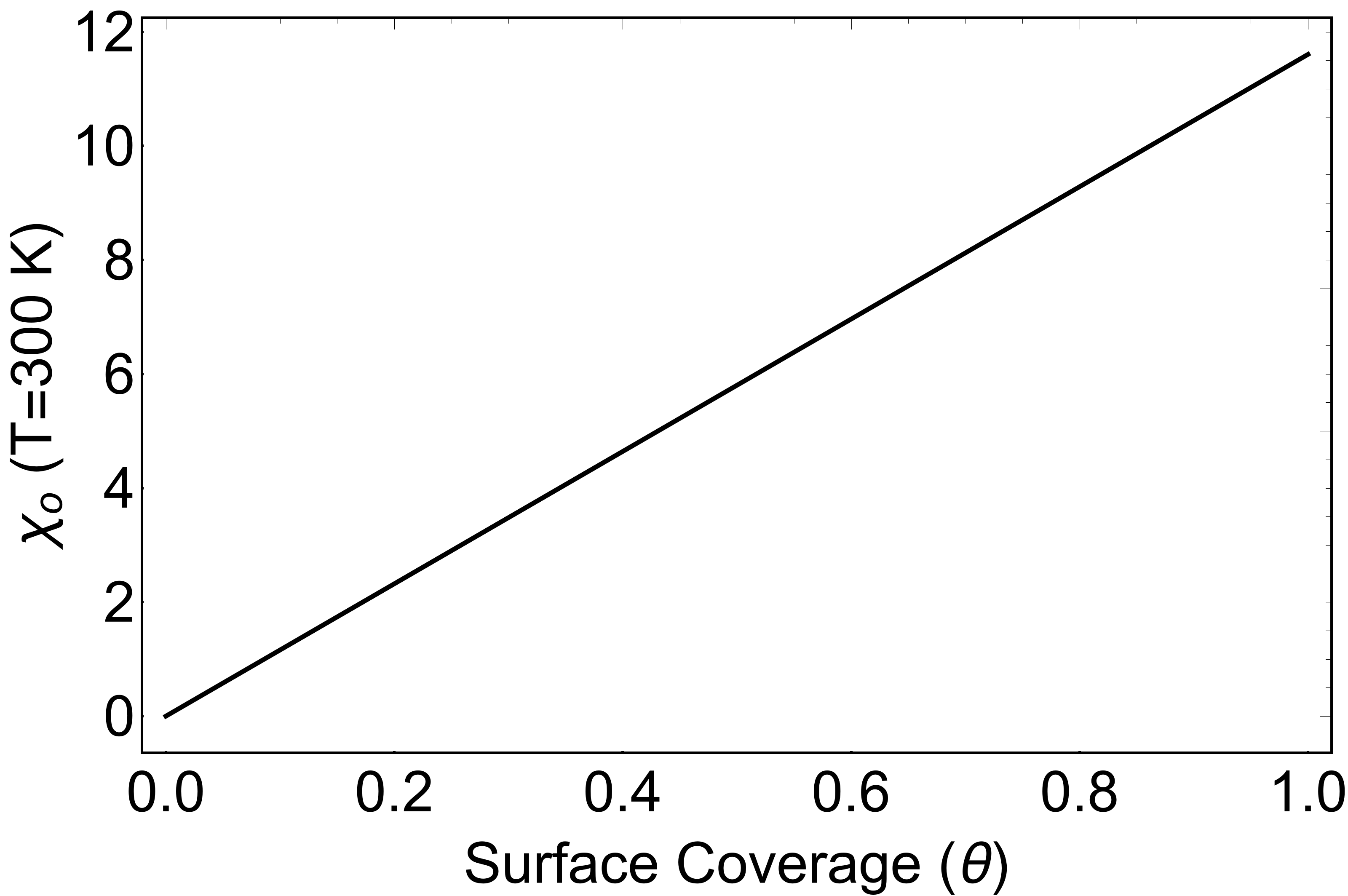}
	\caption{Isotropic susceptibility of a water adlayer physisorbed on fluorine-terminated diamond as a function of surface coverage at 300 K.}
	\label{fig:suscept}
\end{figure}

We now consider the electrostatic response of the dipolar water layer to some perturbing field generated by the charged source, $\vec{E}_p$. Denoting the electric field generated by the C-F surface dipoles as $\vec{E}_s = E_s \hat{z}$, the total electric field can be written as
$$\vec{E} = \vec{E}_s + \vec{E}_p.$$
The polarization of the water layer due to $\vec{E}_p$ can be characterized through a first order Taylor expansion of equation~\eqref{pol} about $\vec{E}_s$ as
\begin{align}
\vec{P}(\vec{E}_s+\vec{E}_p)&=\vec{P}(\vec{E}_s)+\epsilon_0 \overleftrightarrow{\chi}(\vec{E}_s) \cdot \vec{E}_p,
\end{align}
where $\overleftrightarrow{\chi}$ is the linear electric susceptibility tensor given by
\begin{align*}
\epsilon_0 \overleftrightarrow{\chi}_{i,j}(\vec{E}_s) &= \frac{d \vec{P}_i}{d \vec{E}_{p,j}}\Bigg |_{\vec{E}_s} \\
&= \mu \mathcal{N}\left [ \frac{d\mathcal{L}_i}{dE_{p,j}}\Bigg |_{\vec{E}_s}\hat{E_s} +\mathcal{L}(\beta E_s)\frac{d\hat{E}_{s,i}}{dE_{p,j}}\right ] \\
&= \mu \mathcal{N} \Bigg [ \beta \left ( \frac{1}{\beta^2 E_s^2} -\text{csch}^2(\beta E_s) \right)\delta_{i,z} \\
& \ \ \ +\frac{\mathcal{L}(\beta E_s)}{E_s}\delta_{i,j} \Bigg ].
\end{align*}
Consequently, $\overleftrightarrow{\chi}$ may be written as the sum of an isotropic ($\chi_0$) and anisotropic ($\chi_1$) component as
\begin{equation}\label{chi}
\overleftrightarrow{\chi} = \chi_0 \overleftrightarrow{\mathbb{I}} + \chi_1 \hat{z}\hat{z},
\end{equation}
where
\begin{align*}
\epsilon_0\chi_0 &= \frac{P(\beta E_s)}{E_s}, \\
\epsilon_0\chi_1 &= \mu \mathcal{N} \beta \left ( \frac{1}{\beta^2 E_s^2} -\text{csch}^2(\beta E_s) \right)-\epsilon_0 \chi_0.
\end{align*}
Therefore the change in polarization of the adlayer induced by $\vec{E}_p$ is given by
\begin{equation}\label{delP}
\Delta\vec{P} = \epsilon_0 \overleftrightarrow{\chi}\cdot \vec{E}_p = \epsilon_0 \chi_0 \vec{E}_p+\epsilon_0 \chi_1 \vec{E}_{p,z}\hat{z}.
\end{equation}
Taking $E_s\sim0.1$ V/\AA \ derived from \textit{ab-initio} calculations\cite{Mayrhofer2016}, we find that $\chi_1 \ll \chi_0$ at 300~K and therefore the anisotropic term may be neglected. The isotropic susceptibility is presented as a linear function of surface coverage in Figure~\ref{fig:suscept} where we obtain $\chi_0 \approx 12$ for $\theta = 1$. The susceptibility of the water adlayer is therefore relatively large but diminished in respect to liquid water. This is because the ability of the adlayer molecules to re-orientate in response to an electric field is constrained by the surface. In the following section these susceptibilities will be use to model the screening fraction due to both the water adlayer and bulk diamond.

\subsection{Internal diamond}

Debye screening due to internal defects presents a major impediment to diamond electrometry. This form of screening occurs when charges are free to rearrange within a spatial continuum of donors and acceptors\cite{FRS2013}. This causes external fields to decay exponentially within the lattice as they are counteracted by the induced response of the charges. Within diamond, uncontrolled N defects are a common and potent source of Debye screening with a characteristic decay length of 15~nm at low doping concentrations\cite{Broadway2018}. In addition to screening, p-type defects such as boron are detrimental to electrometry as they introduce holes which quench the NV${}^-$ center\cite{Liao2008}. Consequently, only pure diamond is compatible with precision electrometry in which isotropic polarizability is the sole source of internal field decay.

\begin{figure}[]
	\centering
	\includegraphics[width=0.5\textwidth]{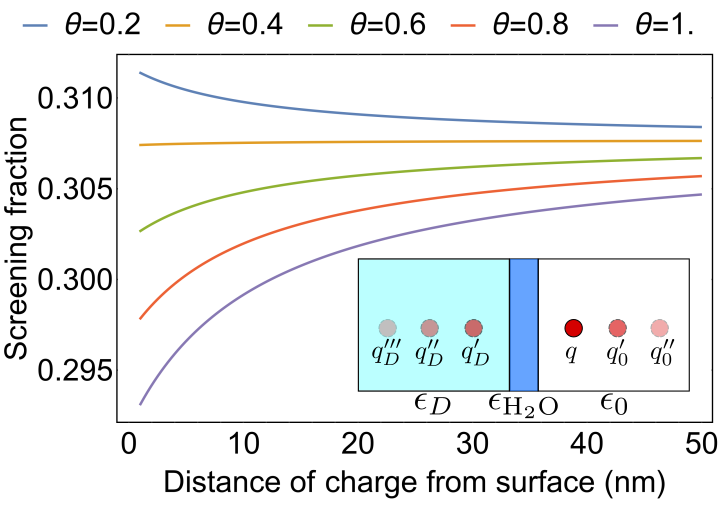}
	\caption{Electric field screening ratio at a depth of 10~nm into diamond for varying water surface coverage. This represents the ratio of the screened field magnitude to an unscreened field magnitude (i.e., within a vacuum) produced by a point charge $(q)$. In this model the water layer is 3~\AA \ wide. The inset depicts the method of images used to determine the screening ratio. Here fictitious `image' charges (dashed $q$s) are introduced to solve Poisson's equation. The magnitude of the charges and their positions are such that they reproduce the electrostatic boundary conditions of the three stacked dielectrics, representing the diamond, water adlayer and external atmosphere.}
	\label{fig:surf_cov}
\end{figure}

Ignoring presently the impact of surface defects, the magnitude of screening due to the adsorbed water layer and pure diamond can be determined analytically. As depicted in the inset of Figure~\ref{fig:surf_cov}, a basic electrometer may be modeled by three stacked planar dielectrics consisting of a thin water layer sandwiched between diamond and air. If an elementary charge is placed above the surface, the electrostatic problem can be solved using the method of images\cite{Barrera1978,Pont2015}. This technique solves Poisson's equation by introducing an infinite series of fictitious `image' charges which reproduce the boundary conditions of the dielectric stack.

In Figure~\ref{fig:surf_cov} we present the electric field screening ratio of a point charge located at varying heights above the surface. This represents the magnitude of the screened field relative to the unscreened field sampled at a depth of 10~nm into the lattice, corresponding to a possible location of a near-surface NV center. A lower estimate for the thickness of the water layer is taken to be 3~\AA, approximately the Van der Waals radius of a single $\text{H}_2\text{O}$ molecule. We obtain screening fractions of $\sim0.3$, independent of surface coverage and the height of the charge above the surface. This is comparable to the screening ratio purely due to the dielectric response of diamond and indicates that adsorption of water vapor is not detrimental to nanoscale electrometry. While a 70\% reduction in field strength appears considerable, this is unavoidable, and minor compared to screening induced by surface charge rearrangement as discussed below.

\subsection{Surface}

Uncontrolled, diamond surface defects present a detrimental source of screening and charge quenching for NV-based electrometers. Recent work has identified a family of primal sp${}^2$ defects universal to all diamond surface terminations\cite{Stacey2019}. These introduce acceptor states into diamond approximately 2.2~eV above the valence band which readily quench the NV${}^-$ charge state at 2.9~eV\cite{Aslam2013b}. Furthermore, partial occupation of these traps results in intense field screening through re-arrangement of surface charge. For fluorine terminated diamond, an sp${}^2$ surface trap density of 4\% has been observed following passivation using SF${}_6$ plasma\cite{Stacey2019}. This corresponds to a trap concentration of roughly $10^{18}$~$\text{m}^{-2}$, comparable to the surface density of free electrons in copper. While the mechanisms for conductivity differ in these materials (i.e., localized defects vs. conduction electrons), such high trap concentrations in diamond effectively render the surface conducting under partial occupation. Surface screening therefore presents the greatest impediment to diamond electrometry and must be addressed in any realistic device.

Uncontrolled, diamond surface defects present a detrimental source of screening and charge quenching for NV-based electrometers. Recent work has identified a family of primal sp${}^2$ defects universal to all diamond surface terminations\cite{Stacey2019}. These introduce acceptor states into diamond approximately 2.2~eV above the valence band which readily quench the NV${}^-$ charge state at 2.9~eV\cite{Aslam2013b}. Furthermore, partial occupation of these traps results in intense field screening through permitting unimpeded rearrangement of surface charge. There is an analogy to a conductor where free carriers may also rearrange to screen an external field. The screening strength is determined by the accessible density of charges that can rearrange. For fluorine terminated diamond, an sp${}^2$ surface trap density of 4\% has been observed following passivation using SF${}_6$ plasma\cite{Stacey2019}, corresponding to a trap concentration of roughly $10^{18}$~$\text{m}^{-2}$. To draw the analogy with a conductor, this is a similar density of free charges as copper. Hence such a diamond surface with a high density of surface traps can be considered as a conductor that screens external electric fields. Surface defects therefore present the greatest impediment to diamond electrometry and must be addressed in any realistic device.

\section{Toy model electrometer}
\label{sec:cap}

A solution to both surface-induced screening and charge instability is to saturate the surface traps using sacrificial donors within the diamond. We propose fabricating a $\delta$-doped layer of $\text{N}_\text{S}$ which introduces a donor level 3.8~eV above the valence band. In the limit that the concentration of $\text{N}_\text{S}$ exceeds that of the sp${}^2$ defects, the Fermi-level will be pinned to the donor level and prevent quenching of the NV${}^-$. This is possible as present doping techniques allow for precision control of the $\delta$-layer height and defect concentrations up to 1000~ppm\cite{Ohno2012a,Chandran2016}. In this section we demonstrate the effectiveness of this idea through an analytical toy model that explores the complex interactions of screening and quenching within a highly coupled system.

\begin{figure}[]
	\centering
	\includegraphics[width=0.45\textwidth]{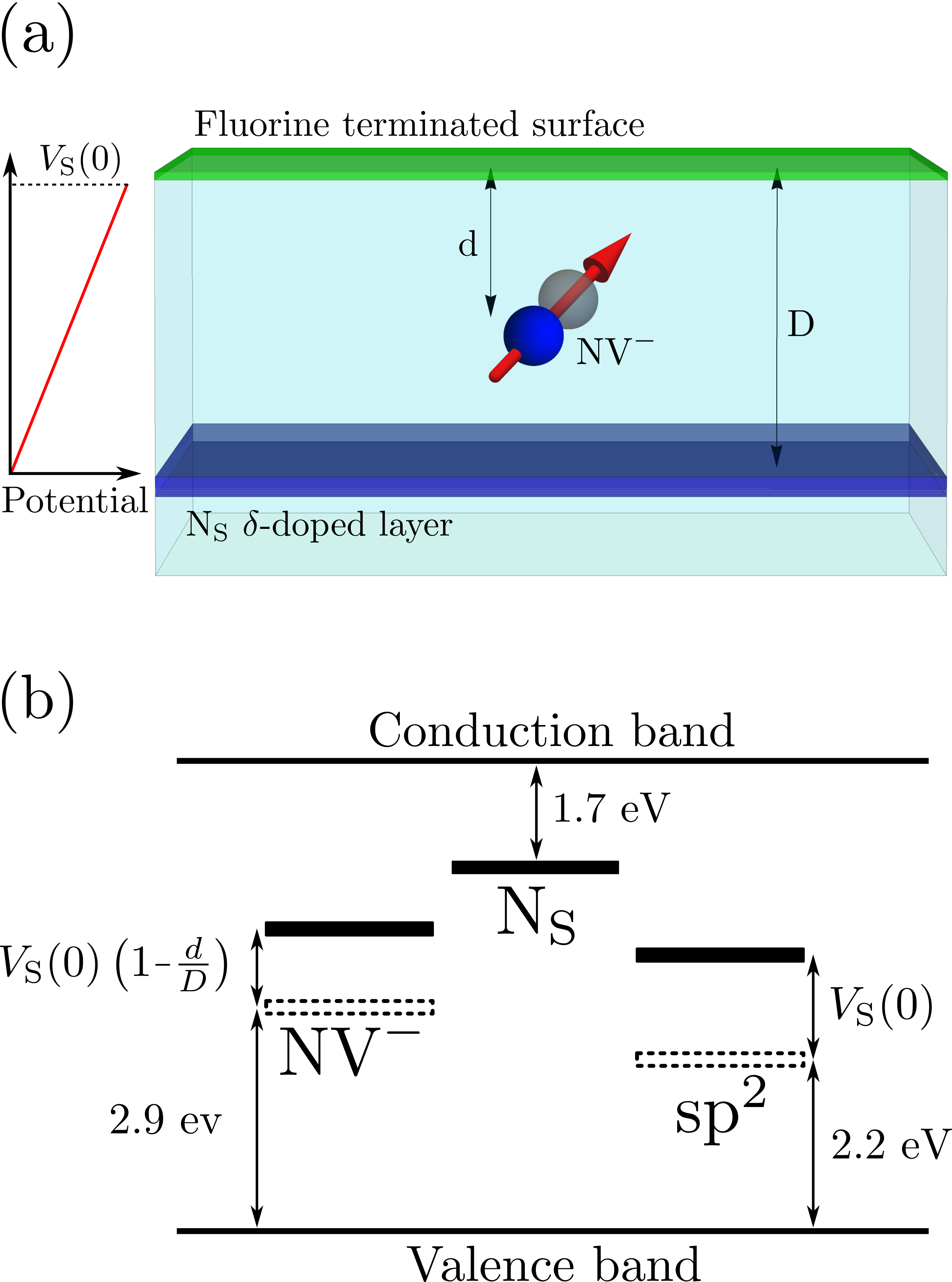}
	\caption{(a) Schematic of the toy model electrometer. An NV center (nitrogen represented as blue sphere, vacancy as grey sphere, and ground-state spin as red arrow) is positioned a depth $d$ below a fluorinated terminated surface. The surface possesses a density $\sigma_T$ of sp${}^2$ charge traps which are preferentially saturated by a grounded $\text{N}_\text{S}$ $\delta$-doped layer positioned at a depth $D$.  This generates a negative charge density at the surface subsequently forming a parallel plate capacitor. A linear potential extends throughout the capacitor with a magnitude $V_S(0)$ at the surface. (b) Energy level diagram of defect states within the diamond electrometer. The acceptor level of the sp${}^2$ charge traps are raised by an energy $V_S(0)$, while that of the NV center is raised by $V_S(0)\left ( 1-\frac{d}{D}\right )$. Charge stability of the NV${}^-$ center requires full occupation of the charge traps and $E_\text{NV} + V_S(0)\left ( 1-\frac{d}{D}\right ) \ll E_{\text{N}}$.}
	\label{fig:toy}
\end{figure}

Consider the schematic for a simplified electrometer presented in Figure~\ref{fig:toy}~(a). The $\delta$-doped layer is positioned at a depth~$D$ below the diamond surface ($z=0$) while the NV spin-probe is placed between them at a depth $d<D$. At electrostatic equilibrium the occupation of the sp${}^2$ defects leads to the accumulation of an isotropic and homogeneous charge density on the surface, $\rho_S$. This charge density subsequently generates a surface potential $V_S(z=0)$ which is related as per
\begin{equation}\label{capacitor}
\rho_S = CV_S(0) = q\sigma_Tf(E_T+qV_S(0)-E_N),
\end{equation}
where $C$ is the device capacitance, $q$ is the electron charge, $\sigma_T$ is the density of surface traps, $f$ is the Fermi-Dirac distribution, $E_N=3.8$~eV is the energy of $\text{N}_\text{S}$ and $E_T\approx2.2$~eV is the energy of an sp${}^2$ surface defect\cite{Stacey2019}. The two layers -- surface and $\delta$-doping -- effectively form a parallel plate capacitor and hence $C=\epsilon_D/D$ where $\epsilon_D=5.7\epsilon_0$ is the dielectric permittivity of diamond. This induces a linear potential between the capacitor plates such that the potential at the NV center is given by
\begin{equation}\label{linearpot}
V_S(d)=V_S(0)\left(1-\frac{d}{D}\right).
\end{equation}

The magnitude of the potential induced at the surface has major ramifications for NV${}^-$ charge stability. Figure~\ref{fig:toy}~(b) depicts the energies of $\text{N}_\text{S}$, NV and sp${}^2$ defects within the simplified electrometer. Upon charging, the sp${}^2$ defect and NV energies are raised by an amount $qV_S(0)$ and $qV_S(d)$ respectively. To avoid transition to NV${}^0$, the condition
\begin{equation}\label{cond1}
E_\text{NV}+qV_S(d)\ll E_N,
\end{equation}
must be maintained. Inserting equation~\eqref{linearpot} into the inequality~\eqref{cond1} places a constraint on the maximum surface potential that prevents quenching, given by
\begin{equation}\label{cond2}
qV_S(0) \ll \frac{E_N-E_\text{NV}}{\left(1-\frac{d}{D}\right)}.
\end{equation}

The surface potential also has major implications for screening. Note that $V_S(0)$ is ultimately limited above by $0\leq qV_S(0)\leq E_N-E_T\approx1.6$~eV. When $qV_S(0)\approx E_N-E_T$ the defect energy is pinned to that of the $\text{N}_\text{S}$ donors and screening effects dominate; equation~\eqref{capacitor} indicates that $\rho_S=q\sigma_T/2$ and hence the surface is effectively conducting. Clearly, electrometry requires that
\begin{equation}\label{ineq1}
qV_S(0) \ll E_N-E_T
\end{equation}
for under such conditions $f(E_T+qV_S(0)-E_N)\approx1$ (the linear regime) and the surface charge cannot reorganize in response to an external electric field. Fortunately, the inequality~\eqref{ineq1} can be determined precisely and the robustness of the linear regime to charge screening can be demonstrated quantitatively.

Consider the screening field induced by a perturbing potential at the diamond surface, $\delta V$. Within the linear regime we have that
\begin{align}
\rho &= q\sigma_T f(E_T + q(V_S(0)+\delta V)-E_N) \nonumber \\
&=q\sigma_T f(\Delta + q\delta V) \nonumber \\
&\approx q \sigma_T \left ( 1 + q\beta e^{-\beta\Delta}\delta V \right ),\label{inducedrho1}
\end{align}
where we have denoted $\Delta \equiv (E_T + qV_S(0)-E_N)$ and expanded to first order in $\delta V$. Suppose that the perturbing potential is due to a point charge $Q$ positioned a height $h$ above the electrometer surface and aligned with the NV center. Then equation~\ref{inducedrho1} indicates that the induced charge density is given by
\begin{align}
\delta \rho(r) &= q^2\sigma_T \beta e^{-\beta\Delta}\delta V \nonumber \\
&=\frac{q^2Q\sigma_T\beta e^{-\beta\Delta}}{4\pi\epsilon_0\sqrt{h^2+r^2}},\label{inducedrho2}
\end{align}
as a function of radial distance from the point charge $r$. This produces a screening field at the NV center given by
\begin{align}
\delta E(d) &= \frac{\hat{z}}{2\epsilon_0}\int_0^\infty \delta\rho(r)\frac{d}{(d^2+r^2)^{3/2}}r\text{d}r \nonumber \\
&= -\frac{q^2Q\sigma_T\beta e^{-\beta\Delta}}{8\pi\epsilon_0^2(d+h)}\hat{z}.
\end{align}
Considering that the field generated by the point charge at a depth $d$ in the absence of screening is
\begin{equation}\label{field}
E(d)=\frac{Q}{4\pi\epsilon_0}\frac{\hat{z}}{(d+h)^2},
\end{equation}
the magnitude of the screening ratio is given by
\begin{equation}\label{rat}
R=\left | \frac{\delta E(d)}{E(d)} \right |=\frac{q^2\sigma_T\beta e^{\beta\Delta}(d+h)}{2\epsilon_0}.
\end{equation}

Equation~\ref{rat} demonstrates a further bound on the maximum surface potential compatible with electrometry. Taking $(d+h)=200$~nm and $\sigma_T = 10^{18}$~$\text{m}^{-2}$ we obtain a screening ratio of $R\approx 10^{-22}e^{\beta qV_S(0)}$. Screening effects will therefore dominate if the surface potential is not sufficiently deep within the linear regime. In this specific example, a ratio $R\leq1\%$ only occurs for $qV_S(0)\leq E_N-E_T - 15 k_B T = 1.2$~eV at room temperature.

To summarize, our toy model has identified the fundamental limitations to precision electrometry. The potential of the surface must be controlled such that all charge traps are saturated while the NV energy is maintained below the Fermi level. In general, one should aim to reduce $V_S(0)$ to avoid the detrimental effects of screening and quenching. Equation~\ref{capacitor} indicates that the surface potential is a function of only two variables; the capacitance and the density of charge traps. For the simplified electrometer design, $V_S(0)\propto C^{-1}$ within the linear regime and therefore $D$ should be minimized. However, $D$ is limited below by physical constraints such as the thickness of the $\delta$-doped layer and its proximity to the surface and NV center. Hence the surface potential is largely dictated by the density of surface traps. Whereas the capacitance can be controlled, surface defects are an undesirable byproduct of diamond surface passivation\cite{Stacey2019}.

\begin{figure}[]
	\centering
	\includegraphics[width=0.45\textwidth]{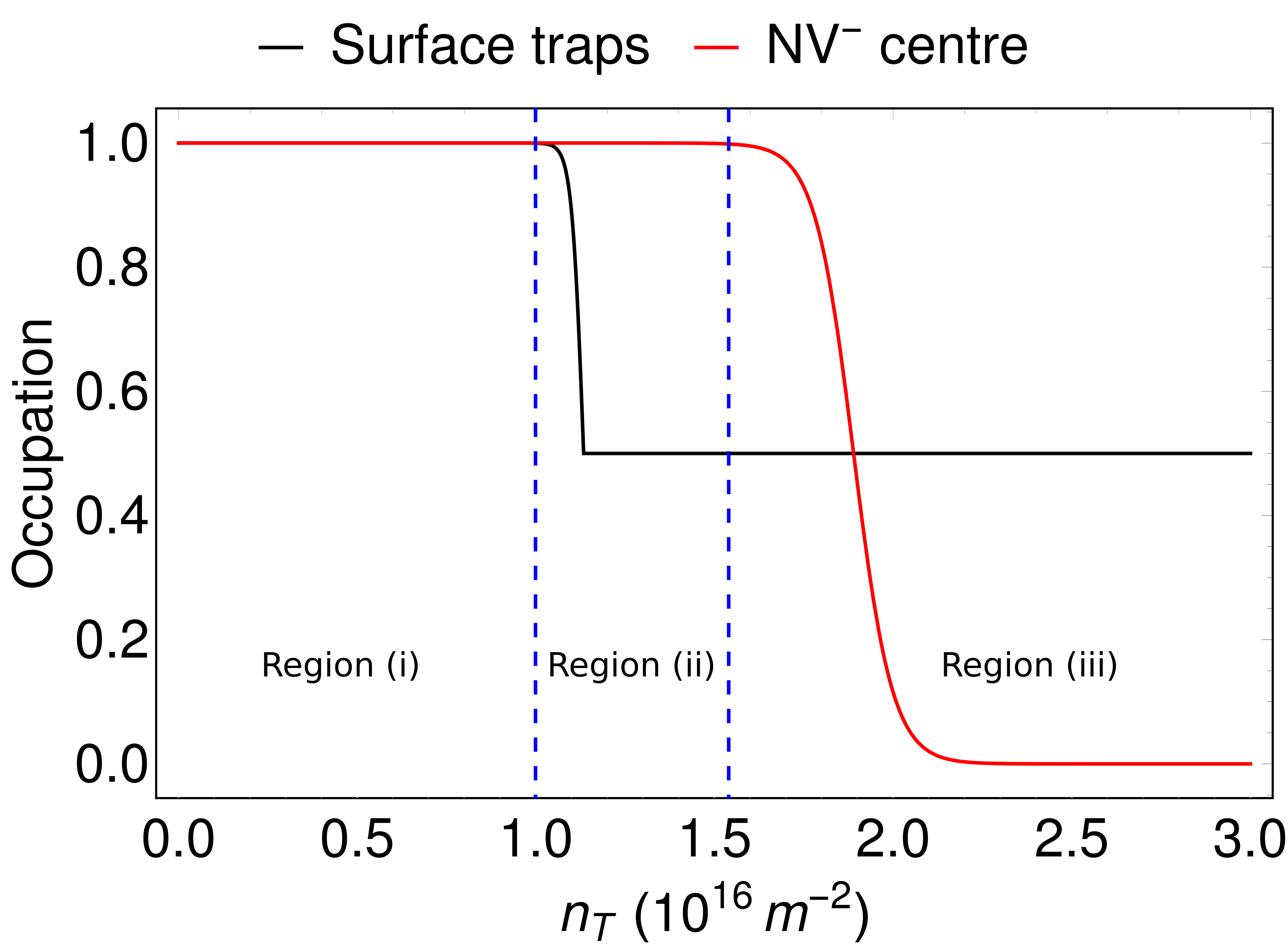}
	\caption{Occupation of surface traps and the NV${}^-$ charge state as a function of surface trap density for the toy model electrometer. Here we assume that $D=100$~nm, $d=60$~nm and $T=300$~K. Three different electrometer operational regions have been distinguished and are discussed in text.}
	\label{fig:toy_occ}
\end{figure}

The performance of a diamond electrometer is strongly dependent on the surface trap density. This relationship is demonstrated in Figure~\ref{fig:toy_occ} where $D=100$~nm, $d=60$~nm and $T=300$~K have been chosen as a realistic example of device parameters. Three different electrometer operating regions can be distinguished. Region~(i) represents values of $\sigma_T$ which correspond to surface voltages within the linear regime. Here electrometry is viable as the charge traps are fully saturated and the NV maintains its negative charge state. Region~(ii) also represents values of $\sigma_T$ for which the NV${}^-$ center remains charge stable. However, the sp${}^2$ defects are only partially occupied and so electrometry is impossible due to surface screening. Similarly, in region~(iii) $\sigma_T$ is so large that the NV center has been quenched.

Consequently, the presence of a $\text{N}_\text{S}$ $\delta$-doped layer can simultaneously maintain NV${}^-$ stability and prevent surface screening within the linear regime. However, the capabilities of the device are fundamentally limited by the density of surface traps. As capacitance is a geometric property, equation~\eqref{capacitor} indicates that this result is universal to any device which employs donors to saturate surface defects. The parallel plate capacitor has one of the greatest capacitances over small length scales and hence Figure~\ref{fig:toy_occ} demonstrates that electrometry is only compatible with surface densities on the order of $\sigma_T \lesssim 10^{16}$~$\text{m}^{-2}$. This is two orders of magnitude lower than defect densities currently observed on fluorine terminated surfaces passivated using SF${}_6$ plasma. However, fluorine is a relatively new surface termination and many alternative passivation techniques exist and continue to be developed\cite{Cui2013,Rietwyk2013,Salvadori2010}. The sp${}^2$ defect yields on these surfaces are yet to be characterized and may well be low enough to permit precision electrometry.

\section{Electrometer design}
\label{sec:design}

The simple electrometer presented in the previous section has several deficiencies which make it impractical for quantum sensing. Fortunately, these can be overcome with a simple modification to the electrometer design. In this section we discuss the shortcomings of the toy-model electrometer and their solutions, culminating in the presentation of an effective and physically realizable device.

\begin{figure}[]
	\centering
	\includegraphics[width=0.5\textwidth]{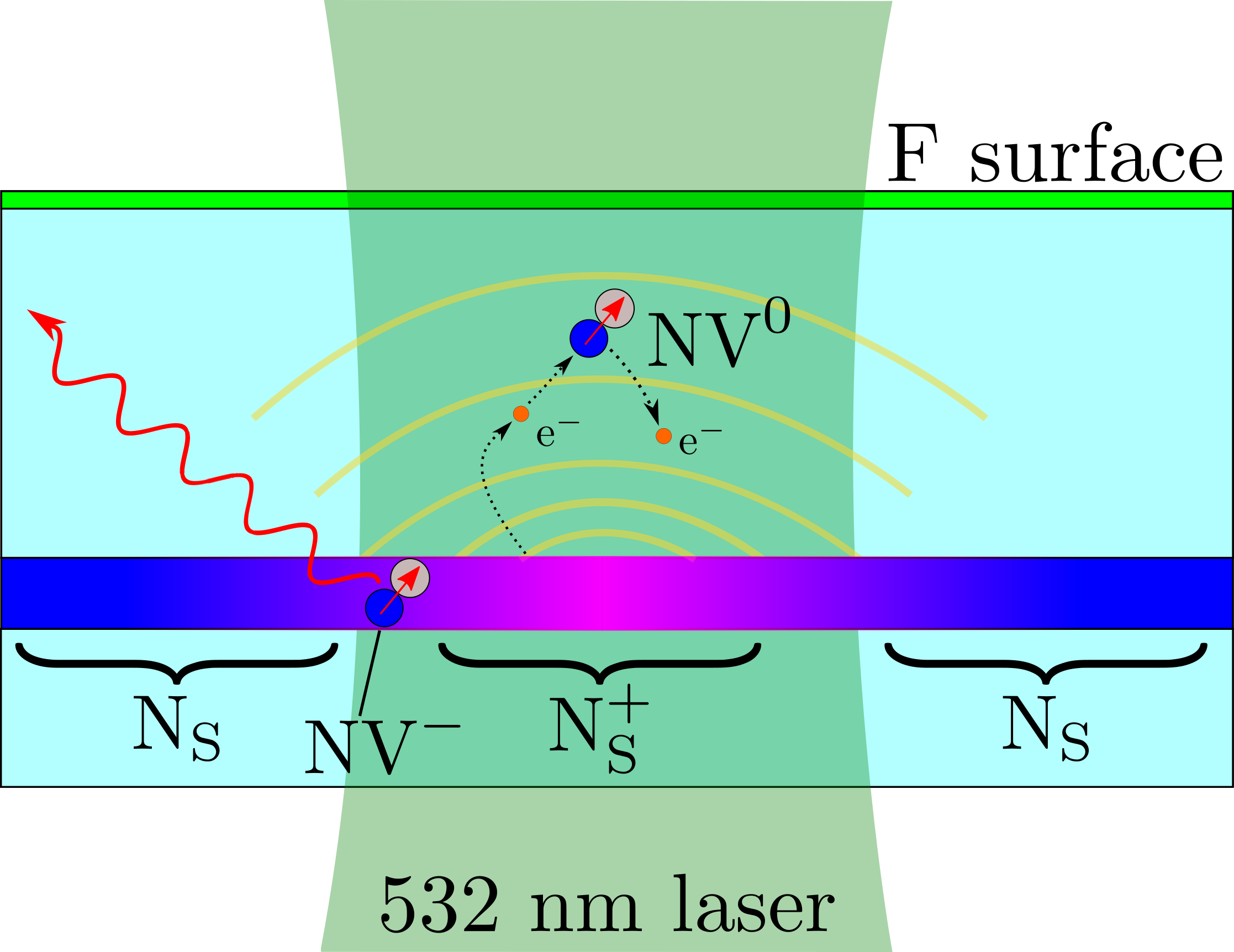}
	\caption{Screening and quenching mechanisms induced by ionization of $\text{N}_\text{S}$ during optical control of the NV center. The 532~nm laser induces a density of $\text{N}_\text{S}^+$ defects within the $\delta$-doped layer (depicted in pink) which subsequently generates a local positive potential (yellow contour lines) and ionizes free electrons into the lattice (orange circles). The local potential has the potential to reduce NV${}^-$ charge stability while the modulating charge density interferes with NV spin dynamics during quantum sensing. There is also a lower risk of NV${}^-$ charge destabilization due to scattering with Auger electrons. Erroneous NV centers present in the doped layer can also introduce background counts during read-out (red photon) which increases measurement contrast leading to lower sensitivity.}
	\label{fig:ionization_2D}
\end{figure}

Initialization and readout of the NV center requires optical control using a 532~nm green laser\cite{Doherty2013,Dolde2011a,Dolde2014a}. As depicted in Figure~\ref{fig:ionization_2D}, performing electrometry with the toy device requires the optical path to pass through the $\delta$-doped $\text{N}_\text{S}$ layer. This introduces several complications. Firstly, $\text{N}_\text{S}$ layers typically contain some density of erroneous NV centers. These are capable of producing background counts during read-out which decrease measurement contrast and lead to lower sensitivity. Secondly, the 532~nm laser ionizes $\text{N}_\text{S}$ defects to form $\text{N}_\text{S}^+$ which modulates the charge density within the $\delta$-doped layer\cite{Iakoubovskii2000}. During the optical steady state this results in a local positive potential which reduces charge stability of the $\text{NV}^-$ center and further diminishes signal contrast. Moreover, following initialization of the NV${}^-$ spin-state the laser is deactivated and the induced charge relaxes back to the equilibrium it adopted before the laser was turned on. If this relaxation time is slow ($\approx100$~ns) compared to the NV sensing period ($> 1$~$\mu$s)\cite{Dolde2011a,Dolde2014a}, this relaxation will influence the sensing measurement. Ionization of $\text{N}_\text{S}$ also introduces free charge carriers into the lattice, and hence there also exists a low probability of Auger electrons scattering against the NV${}^-$ and causing quenching. Finally, the $\delta$-doped $\text{N}_\text{S}$ layer acts as a spin-bath which can cause decoherence of the NV spin at close proximities\cite{Balasubramanian2009}.

\begin{figure}[]
	\centering
	\includegraphics[width=0.5\textwidth]{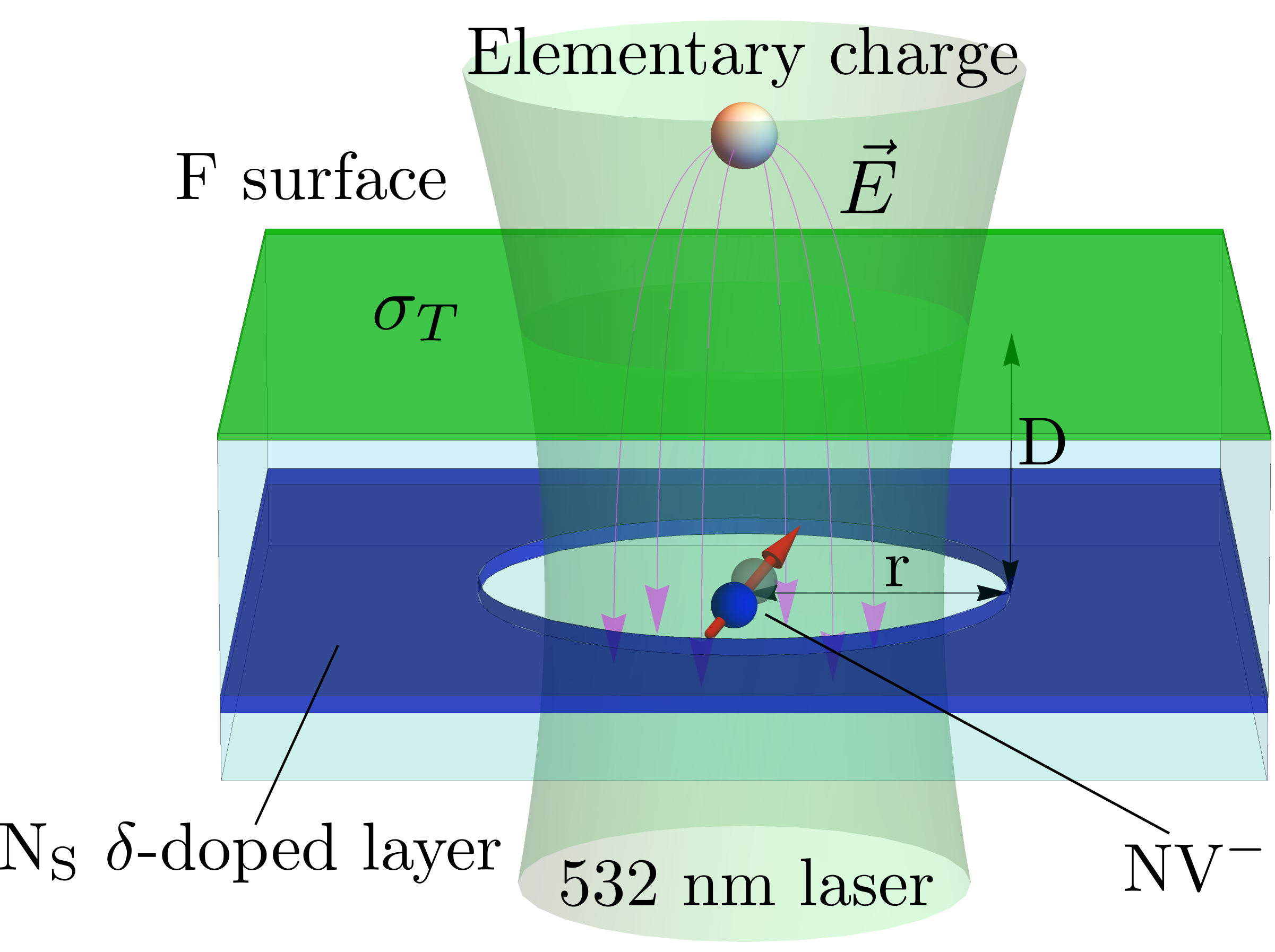}
	\caption{Schematic of a diamond-based electrometer for sub-nanometer resolution charge imaging at ambient temperatures. A shallow NV${}^-$ spin probe detects the electric field from an elementary charge positioned above the surface through optically detected magnetic resonance. Primal sp${}^2$ defects present on the surface with density $\sigma_T$ act as electron traps, quenching the NV${}^-$ charge state and causing detrimental screening when partially occupied. This is systematically controlled by saturating the charge traps using a sacrificial layer of $\delta$-doped $\text{N}_\text{S}$ positioned a depth $D$ into the substrate. A $\text{N}_\text{S}$-deficit disk of radius $r$ allows for optical initialization and readout of the NV${}^-$ spin without ionizing mobile charge carriers from the $\delta$-doped layer.}
	\label{fig:device}
\end{figure}

One possible solution to these issues is to spatially separate the $\text{N}_\text{S}$ layer and the NV center. The optical focus can then be maintained on the NV while the $\delta$-doping is subject to a negligible laser intensity. We estimate this would require a separation distance of approximately $0.5$~$\mu$m. However, this severely limits the sensitivity of the spin-probe to external charges as equation~\eqref{cond2} necessitates that $D\gtrsim1$~$\mu$m and therefore $d\gtrsim0.5$~$\mu$m to maintain $\text{NV}^-$ charge stability. A more practical solution is to introduce of a hole within the $\delta$-doped layer. Consider the schematic presented in Figure~\ref{fig:device} in which a disk of pure diamond has been fabricated around the NV center. This hole permits optical access to the spin probe while simultaneously minimizing the number of ionized $\text{N}_\text{S}$ defects. Furthermore, the hole reduces the probability of optically addressing multiple NV centers and so increases the yield of electrometer fabrication.

The capabilities of this realistic electrometer design for elementary charge detection were investigated using COMSOL Multiphysics software. The potential at the NV center and surface as well as the occupation of charge traps were simulated as a function of the device parameters; the NV depth $d$, $\text{N}_\text{S}$ layer depth $D$ and hole radius $r$. This was achieved by solving Poisson's equation self-consistently for a surface charge density given by equation~\ref{capacitor} and assuming a grounded $\delta$-doped layer. The device parameters were then optimized to identify the greatest possible trap density compatible with precision electrometry. As discussed in Section~\ref{sec:cap}, these criterion are charge stability of the NV ($qV_S(d) \ll E_N-E_{NV}=0.9$~eV) and a surface potential which leads to less than 1\% surface screening as per equation~\eqref{rat}.

\begin{figure}[]
	\centering
	\includegraphics[width=0.5\textwidth]{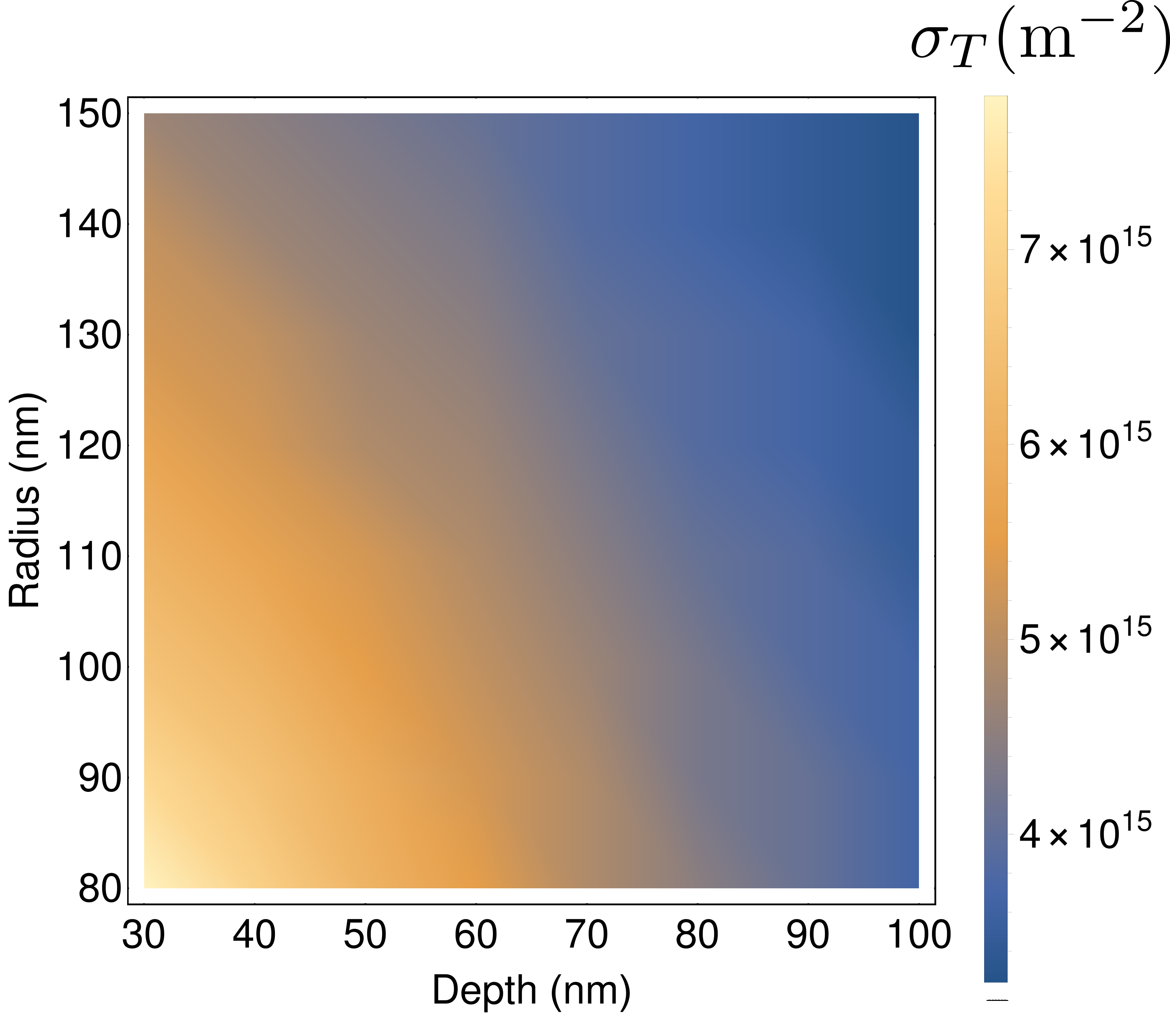}
	\caption{Maximum viable surface trap density ($\sigma_T$) as a function of $\delta$-doped $\text{N}_\text{S}$ layer depth and hole radius for the electrometer design presented in Figure~\ref{fig:device}. Trap densities in excess of those presented here result in an electric field screening ratio in excess of 1\%.}
	\label{fig:maxNt}
\end{figure}

Figure~\ref{fig:maxNt} presents the largest viable surface trap density as a function of the $\delta$-doping depth and hole radius. Trap densities greater than those presented lead to field screening in excess of 1\%. For the parameters sampled here ($30$~$\text{nm} < D < 100$~nm and $80$~$\text{nm} < r < 150$~nm) we find that electrometry is feasible for sp${}^2$ surface densities within the regime of $10^{15}$~$\text{m}^{-2}$ and that no charge quenching occurs for NV centers level with the $\delta$-doped layer. Devices with smaller radii and $\delta$-doping depth possess a greater capacitance and are therefore compatible with larger defect densities. Note that the optical spot size is diffraction limited by $\approx200$~nm, and hence some degree of $\text{N}_\text{S}$ ionization will occur for hole radii less than 100~nm. The impact of the induced positive charge density for NV quantum sensing is unknown and left as an avenue for future work. If the effects are significantly detrimental, Figure~\ref{fig:maxNt} indicates that the hole size can simply be increased beyond the optical spot size without drastically reducing the achievable sp${}^2$ defect concentrations.

\section{Conclusion}
\label{sec:conc}
Sub-nanometer resolution electrometry of elementary charges under ambient conditions would allow for investigation of diverse electrical phenomenon ranging from biological systems to fundamental physics. The NV center is the only known system capable of such a feat, however measurements are currently limited to charges internal to diamond. In this paper we have applied theoretical modeling to conclusively demonstrate that external charge detection is not yet feasible due to field screening. While screening due to the atmosphere and internal defects can be mitigated using fluorine-passivated and ultra-pure diamond, electrometry is ultimately frustrated by charge rearrangement amongst surface defects.

We have proposed a solution to surface screening through introduction of a sacrificial $\text{N}_\text{S}$ $\delta$-doped layer. Fabrication of a $\text{N}_\text{S}$ deficit hole surrounding the NV center allows for optical access to the spin probe while minimizing the read-out of erroneous NV centers and ionization of free charges. This electrometry device is technologically feasible and computational simulations have demonstrated that it can successfully mitigate screening effects for surface trap densities up to $\approx10^{16}$~$\text{m}^{-2}$. Although this is two orders of magnitude below currently observed sp${}^2$ defect densities on fluorine-terminated diamond, the outcome of this work is a clear pathway towards nanoscale imaging of external charges at ambient conditions. Electrometry cannot be achieved until surface passivation technologies realize lower defect concentrations on fluorine-terminated diamond.

\section*{Acknowledgments}
We acknowledge funding from the Australian Research Council (DP170102735). We thank Patrick Maletinsky for providing insightful feedback on this manuscript.

\bibliography{screeningOberg_final}

\begin{thebibliography}{62}%
\makeatletter
\providecommand \@ifxundefined [1]{%
 \@ifx{#1\undefined}
}%
\providecommand \@ifnum [1]{%
 \ifnum #1\expandafter \@firstoftwo
 \else \expandafter \@secondoftwo
 \fi
}%
\providecommand \@ifx [1]{%
 \ifx #1\expandafter \@firstoftwo
 \else \expandafter \@secondoftwo
 \fi
}%
\providecommand \natexlab [1]{#1}%
\providecommand \enquote  [1]{``#1''}%
\providecommand \bibnamefont  [1]{#1}%
\providecommand \bibfnamefont [1]{#1}%
\providecommand \citenamefont [1]{#1}%
\providecommand \href@noop [0]{\@secondoftwo}%
\providecommand \href [0]{\begingroup \@sanitize@url \@href}%
\providecommand \@href[1]{\@@startlink{#1}\@@href}%
\providecommand \@@href[1]{\endgroup#1\@@endlink}%
\providecommand \@sanitize@url [0]{\catcode `\\12\catcode `\$12\catcode
  `\&12\catcode `\#12\catcode `\^12\catcode `\_12\catcode `\%12\relax}%
\providecommand \@@startlink[1]{}%
\providecommand \@@endlink[0]{}%
\providecommand \url  [0]{\begingroup\@sanitize@url \@url }%
\providecommand \@url [1]{\endgroup\@href {#1}{\urlprefix }}%
\providecommand \urlprefix  [0]{URL }%
\providecommand \Eprint [0]{\href }%
\providecommand \doibase [0]{https://doi.org/}%
\providecommand \selectlanguage [0]{\@gobble}%
\providecommand \bibinfo  [0]{\@secondoftwo}%
\providecommand \bibfield  [0]{\@secondoftwo}%
\providecommand \translation [1]{[#1]}%
\providecommand \BibitemOpen [0]{}%
\providecommand \bibitemStop [0]{}%
\providecommand \bibitemNoStop [0]{.\EOS\space}%
\providecommand \EOS [0]{\spacefactor3000\relax}%
\providecommand \BibitemShut  [1]{\csname bibitem#1\endcsname}%
\let\auto@bib@innerbib\@empty
\bibitem [{\citenamefont {Cui}\ \emph {et~al.}(2001)\citenamefont {Cui},
  \citenamefont {Wei}, \citenamefont {Park},\ and\ \citenamefont
  {Lieber}}]{Cui2001}%
  \BibitemOpen
  \bibfield  {author} {\bibinfo {author} {\bibfnamefont {Y.}~\bibnamefont
  {Cui}}, \bibinfo {author} {\bibfnamefont {Q.}~\bibnamefont {Wei}}, \bibinfo
  {author} {\bibfnamefont {H.}~\bibnamefont {Park}},\ and\ \bibinfo {author}
  {\bibfnamefont {C.~M.}\ \bibnamefont {Lieber}},\ }\bibfield  {title}
  {\bibinfo {title} {{Nanowire nanosensors for highly sensitive and selective
  detection of biological and chemical species}},\ }\href
  {https://doi.org/10.1126/science.1062711} {\bibfield  {journal} {\bibinfo
  {journal} {Science}\ }\textbf {\bibinfo {volume} {293}},\ \bibinfo {pages}
  {1289} (\bibinfo {year} {2001})}\BibitemShut {NoStop}%
\bibitem [{\citenamefont {Patolsky}\ \emph {et~al.}(2006)\citenamefont
  {Patolsky}, \citenamefont {Zheng},\ and\ \citenamefont
  {Lieber}}]{Patolsky2006}%
  \BibitemOpen
  \bibfield  {author} {\bibinfo {author} {\bibfnamefont {F.}~\bibnamefont
  {Patolsky}}, \bibinfo {author} {\bibfnamefont {G.}~\bibnamefont {Zheng}},\
  and\ \bibinfo {author} {\bibfnamefont {C.~M.}\ \bibnamefont {Lieber}},\
  }\bibfield  {title} {\bibinfo {title} {{Fabrication of silicon nanowire
  devices for ultrasensitive, label-free, real-time detection of biological and
  chemical species}},\ }\href {https://doi.org/10.1038/nprot.2006.227}
  {\bibfield  {journal} {\bibinfo  {journal} {Nature Protocols}\ }\textbf
  {\bibinfo {volume} {1}},\ \bibinfo {pages} {1711} (\bibinfo {year}
  {2006})}\BibitemShut {NoStop}%
\bibitem [{\citenamefont {Eizerman}\ \emph {et~al.}(2004)\citenamefont
  {Eizerman}, \citenamefont {Hanson}, \citenamefont {{Van Beveren}},
  \citenamefont {Witkamp}, \citenamefont {Vandersypen},\ and\ \citenamefont
  {Kouwenhoven}}]{Eizerman2004}%
  \BibitemOpen
  \bibfield  {author} {\bibinfo {author} {\bibfnamefont {J.~M.}\ \bibnamefont
  {Eizerman}}, \bibinfo {author} {\bibfnamefont {R.}~\bibnamefont {Hanson}},
  \bibinfo {author} {\bibfnamefont {L.~H.}\ \bibnamefont {{Van Beveren}}},
  \bibinfo {author} {\bibfnamefont {B.}~\bibnamefont {Witkamp}}, \bibinfo
  {author} {\bibfnamefont {L.~M.}\ \bibnamefont {Vandersypen}},\ and\ \bibinfo
  {author} {\bibfnamefont {L.~P.}\ \bibnamefont {Kouwenhoven}},\ }\bibfield
  {title} {\bibinfo {title} {{Single-shot read-out of an individual electron
  spin in a quantum dot}},\ }\href {https://doi.org/10.1038/nature02693}
  {\bibfield  {journal} {\bibinfo  {journal} {Nature}\ }\textbf {\bibinfo
  {volume} {430}},\ \bibinfo {pages} {431} (\bibinfo {year}
  {2004})}\BibitemShut {NoStop}%
\bibitem [{\citenamefont {Martin}\ \emph {et~al.}(2008)\citenamefont {Martin},
  \citenamefont {Akerman}, \citenamefont {Ulbricht}, \citenamefont {Lohmann},
  \citenamefont {Smet}, \citenamefont {{Von Klitzing}},\ and\ \citenamefont
  {Yacoby}}]{Martin2008}%
  \BibitemOpen
  \bibfield  {author} {\bibinfo {author} {\bibfnamefont {J.}~\bibnamefont
  {Martin}}, \bibinfo {author} {\bibfnamefont {N.}~\bibnamefont {Akerman}},
  \bibinfo {author} {\bibfnamefont {G.}~\bibnamefont {Ulbricht}}, \bibinfo
  {author} {\bibfnamefont {T.}~\bibnamefont {Lohmann}}, \bibinfo {author}
  {\bibfnamefont {J.~H.}\ \bibnamefont {Smet}}, \bibinfo {author}
  {\bibfnamefont {K.}~\bibnamefont {{Von Klitzing}}},\ and\ \bibinfo {author}
  {\bibfnamefont {A.}~\bibnamefont {Yacoby}},\ }\bibfield  {title} {\bibinfo
  {title} {{Observation of electron-hole puddles in graphene using a scanning
  single-electron transistor}},\ }\href {https://doi.org/10.1038/nphys781}
  {\bibfield  {journal} {\bibinfo  {journal} {Nature Physics}\ }\textbf
  {\bibinfo {volume} {4}},\ \bibinfo {pages} {144} (\bibinfo {year}
  {2008})}\BibitemShut {NoStop}%
\bibitem [{\citenamefont {Yoo}\ \emph {et~al.}(1997)\citenamefont {Yoo},
  \citenamefont {Fulton}, \citenamefont {Hess}, \citenamefont {Willett},
  \citenamefont {Dunkleberger}, \citenamefont {Chichester}, \citenamefont
  {Pfeiffer},\ and\ \citenamefont {West}}]{Yoo1997}%
  \BibitemOpen
  \bibfield  {author} {\bibinfo {author} {\bibfnamefont {M.~J.}\ \bibnamefont
  {Yoo}}, \bibinfo {author} {\bibfnamefont {T.~A.}\ \bibnamefont {Fulton}},
  \bibinfo {author} {\bibfnamefont {H.~F.}\ \bibnamefont {Hess}}, \bibinfo
  {author} {\bibfnamefont {R.~L.}\ \bibnamefont {Willett}}, \bibinfo {author}
  {\bibfnamefont {L.~N.}\ \bibnamefont {Dunkleberger}}, \bibinfo {author}
  {\bibfnamefont {R.~J.}\ \bibnamefont {Chichester}}, \bibinfo {author}
  {\bibfnamefont {L.~N.}\ \bibnamefont {Pfeiffer}},\ and\ \bibinfo {author}
  {\bibfnamefont {K.~W.}\ \bibnamefont {West}},\ }\bibfield  {title} {\bibinfo
  {title} {{Scanning single-electron transistor microscopy: Imaging individual
  charges}},\ }\href {https://doi.org/10.1126/science.276.5312.579} {\bibfield
  {journal} {\bibinfo  {journal} {Science}\ }\textbf {\bibinfo {volume}
  {276}},\ \bibinfo {pages} {579} (\bibinfo {year} {1997})}\BibitemShut
  {NoStop}%
\bibitem [{\citenamefont {Henning}\ \emph {et~al.}(1995)\citenamefont
  {Henning}, \citenamefont {Hochwitz}, \citenamefont {Slinkman}, \citenamefont
  {Never}, \citenamefont {Hoffmann}, \citenamefont {Kaszuba},\ and\
  \citenamefont {Daghlian}}]{Henning1995}%
  \BibitemOpen
  \bibfield  {author} {\bibinfo {author} {\bibfnamefont {A.~K.}\ \bibnamefont
  {Henning}}, \bibinfo {author} {\bibfnamefont {T.}~\bibnamefont {Hochwitz}},
  \bibinfo {author} {\bibfnamefont {J.}~\bibnamefont {Slinkman}}, \bibinfo
  {author} {\bibfnamefont {J.}~\bibnamefont {Never}}, \bibinfo {author}
  {\bibfnamefont {S.}~\bibnamefont {Hoffmann}}, \bibinfo {author}
  {\bibfnamefont {P.}~\bibnamefont {Kaszuba}},\ and\ \bibinfo {author}
  {\bibfnamefont {C.}~\bibnamefont {Daghlian}},\ }\bibfield  {title} {\bibinfo
  {title} {{Two-dimensional surface dopant profiling in silicon using scanning
  Kelvin probe microscopy}},\ }\href {https://doi.org/10.1063/1.358819}
  {\bibfield  {journal} {\bibinfo  {journal} {Journal of Applied Physics}\
  }\textbf {\bibinfo {volume} {77}},\ \bibinfo {pages} {1888} (\bibinfo {year}
  {1995})}\BibitemShut {NoStop}%
\bibitem [{\citenamefont {Williams}\ \emph {et~al.}(1989)\citenamefont
  {Williams}, \citenamefont {Slinkman}, \citenamefont {Hough},\ and\
  \citenamefont {Wickramasinghe}}]{Williams1989}%
  \BibitemOpen
  \bibfield  {author} {\bibinfo {author} {\bibfnamefont {C.~C.}\ \bibnamefont
  {Williams}}, \bibinfo {author} {\bibfnamefont {J.}~\bibnamefont {Slinkman}},
  \bibinfo {author} {\bibfnamefont {W.~P.}\ \bibnamefont {Hough}},\ and\
  \bibinfo {author} {\bibfnamefont {H.~K.}\ \bibnamefont {Wickramasinghe}},\
  }\bibfield  {title} {\bibinfo {title} {{Lateral dopant profiling with 200 nm
  resolution by scanning capacitance microscopy}},\ }\href
  {https://doi.org/10.1063/1.102312} {\bibfield  {journal} {\bibinfo  {journal}
  {Applied Physics Letters}\ }\textbf {\bibinfo {volume} {55}},\ \bibinfo
  {pages} {1662} (\bibinfo {year} {1989})}\BibitemShut {NoStop}%
\bibitem [{\citenamefont {Devoret}\ and\ \citenamefont
  {Schoelkopf}(2000)}]{Devoret2000}%
  \BibitemOpen
  \bibfield  {author} {\bibinfo {author} {\bibfnamefont {M.~H.}\ \bibnamefont
  {Devoret}}\ and\ \bibinfo {author} {\bibfnamefont {R.~J.}\ \bibnamefont
  {Schoelkopf}},\ }\bibfield  {title} {\bibinfo {title} {{Amplifying quantum
  signals with the single-electron transistor}},\ }\href
  {https://doi.org/10.1038/35023253} {\bibfield  {journal} {\bibinfo  {journal}
  {Nature}\ }\textbf {\bibinfo {volume} {406}},\ \bibinfo {pages} {1039}
  (\bibinfo {year} {2000})}\BibitemShut {NoStop}%
\bibitem [{\citenamefont {Sch{\"{o}}nenberger}\ and\ \citenamefont
  {Alvarado}(1990)}]{Schonenberger1990}%
  \BibitemOpen
  \bibfield  {author} {\bibinfo {author} {\bibfnamefont {C.}~\bibnamefont
  {Sch{\"{o}}nenberger}}\ and\ \bibinfo {author} {\bibfnamefont {S.~F.}\
  \bibnamefont {Alvarado}},\ }\bibfield  {title} {\bibinfo {title}
  {{Observation of single charge carriers by force microscopy}},\ }\href
  {https://doi.org/10.1103/PhysRevLett.65.3162} {\bibfield  {journal} {\bibinfo
   {journal} {Physical Review Letters}\ }\textbf {\bibinfo {volume} {65}},\
  \bibinfo {pages} {3162} (\bibinfo {year} {1990})}\BibitemShut {NoStop}%
\bibitem [{\citenamefont {Martin}\ \emph {et~al.}(1988)\citenamefont {Martin},
  \citenamefont {Abraham},\ and\ \citenamefont {Wickramasinghe}}]{Martin1988}%
  \BibitemOpen
  \bibfield  {author} {\bibinfo {author} {\bibfnamefont {Y.}~\bibnamefont
  {Martin}}, \bibinfo {author} {\bibfnamefont {D.~W.}\ \bibnamefont
  {Abraham}},\ and\ \bibinfo {author} {\bibfnamefont {H.~K.}\ \bibnamefont
  {Wickramasinghe}},\ }\bibfield  {title} {\bibinfo {title} {{High-resolution
  capacitance measurement and potentiometry by force microscopy}},\ }\href
  {https://doi.org/10.1063/1.99224} {\bibfield  {journal} {\bibinfo  {journal}
  {Applied Physics Letters}\ }\textbf {\bibinfo {volume} {52}},\ \bibinfo
  {pages} {1103} (\bibinfo {year} {1988})}\BibitemShut {NoStop}%
\bibitem [{\citenamefont {Cleland}\ and\ \citenamefont
  {Roukes}(1998)}]{Cleland1998}%
  \BibitemOpen
  \bibfield  {author} {\bibinfo {author} {\bibfnamefont {A.~N.}\ \bibnamefont
  {Cleland}}\ and\ \bibinfo {author} {\bibfnamefont {M.~L.}\ \bibnamefont
  {Roukes}},\ }\bibfield  {title} {\bibinfo {title} {{A nanometre-scale
  mechanical electrometer}},\ }\href {https://doi.org/10.1038/32373} {\bibfield
   {journal} {\bibinfo  {journal} {Nature}\ }\textbf {\bibinfo {volume}
  {392}},\ \bibinfo {pages} {160} (\bibinfo {year} {1998})}\BibitemShut
  {NoStop}%
\bibitem [{\citenamefont {Bunch}\ \emph {et~al.}(2007)\citenamefont {Bunch},
  \citenamefont {{Van Der Zande}}, \citenamefont {Verbridge}, \citenamefont
  {Frank}, \citenamefont {Tanenbaum}, \citenamefont {Parpia}, \citenamefont
  {Craighead},\ and\ \citenamefont {McEuen}}]{Bunch2007}%
  \BibitemOpen
  \bibfield  {author} {\bibinfo {author} {\bibfnamefont {J.~S.}\ \bibnamefont
  {Bunch}}, \bibinfo {author} {\bibfnamefont {A.~M.}\ \bibnamefont {{Van Der
  Zande}}}, \bibinfo {author} {\bibfnamefont {S.~S.}\ \bibnamefont
  {Verbridge}}, \bibinfo {author} {\bibfnamefont {I.~W.}\ \bibnamefont
  {Frank}}, \bibinfo {author} {\bibfnamefont {D.~M.}\ \bibnamefont
  {Tanenbaum}}, \bibinfo {author} {\bibfnamefont {J.~M.}\ \bibnamefont
  {Parpia}}, \bibinfo {author} {\bibfnamefont {H.~G.}\ \bibnamefont
  {Craighead}},\ and\ \bibinfo {author} {\bibfnamefont {P.~L.}\ \bibnamefont
  {McEuen}},\ }\bibfield  {title} {\bibinfo {title} {{Electromechanical
  resonators from graphene sheets}},\ }\href
  {https://doi.org/10.1126/science.1136836} {\bibfield  {journal} {\bibinfo
  {journal} {Science}\ }\textbf {\bibinfo {volume} {315}},\ \bibinfo {pages}
  {490} (\bibinfo {year} {2007})}\BibitemShut {NoStop}%
\bibitem [{\citenamefont {Salfi}\ \emph {et~al.}(2010)\citenamefont {Salfi},
  \citenamefont {Savelyev}, \citenamefont {Blumin}, \citenamefont {Nair},\ and\
  \citenamefont {Ruda}}]{Salfi2010}%
  \BibitemOpen
  \bibfield  {author} {\bibinfo {author} {\bibfnamefont {J.}~\bibnamefont
  {Salfi}}, \bibinfo {author} {\bibfnamefont {I.~G.}\ \bibnamefont {Savelyev}},
  \bibinfo {author} {\bibfnamefont {M.}~\bibnamefont {Blumin}}, \bibinfo
  {author} {\bibfnamefont {S.~V.}\ \bibnamefont {Nair}},\ and\ \bibinfo
  {author} {\bibfnamefont {H.~E.}\ \bibnamefont {Ruda}},\ }\bibfield  {title}
  {\bibinfo {title} {{Direct observation of single-charge-detection capability
  of nanowire field-effect transistors}},\ }\href
  {https://doi.org/10.1038/nnano.2010.180} {\bibfield  {journal} {\bibinfo
  {journal} {Nature Nanotechnology}\ }\textbf {\bibinfo {volume} {5}},\
  \bibinfo {pages} {737} (\bibinfo {year} {2010})}\BibitemShut {NoStop}%
\bibitem [{\citenamefont {Lee}\ \emph {et~al.}(2008)\citenamefont {Lee},
  \citenamefont {Zhu},\ and\ \citenamefont {Seshia}}]{Lee2008}%
  \BibitemOpen
  \bibfield  {author} {\bibinfo {author} {\bibfnamefont {J.}~\bibnamefont
  {Lee}}, \bibinfo {author} {\bibfnamefont {Y.}~\bibnamefont {Zhu}},\ and\
  \bibinfo {author} {\bibfnamefont {A.}~\bibnamefont {Seshia}},\ }\bibfield
  {title} {\bibinfo {title} {{Room temperature electrometry with SUB-10
  electron charge resolution}},\ }\bibfield  {journal} {\bibinfo  {journal}
  {Journal of Micromechanics and Microengineering}\ }\textbf {\bibinfo {volume}
  {18}},\ \href {https://doi.org/10.1088/0960-1317/18/2/025033}
  {10.1088/0960-1317/18/2/025033} (\bibinfo {year} {2008})\BibitemShut
  {NoStop}%
\bibitem [{\citenamefont {Barry}\ \emph {et~al.}(2016)\citenamefont {Barry},
  \citenamefont {Turner}, \citenamefont {Schloss}, \citenamefont {Glenn},
  \citenamefont {Song}, \citenamefont {Lukin}, \citenamefont {Park},\ and\
  \citenamefont {Walsworth}}]{Barry2016}%
  \BibitemOpen
  \bibfield  {author} {\bibinfo {author} {\bibfnamefont {J.~F.}\ \bibnamefont
  {Barry}}, \bibinfo {author} {\bibfnamefont {M.~J.}\ \bibnamefont {Turner}},
  \bibinfo {author} {\bibfnamefont {J.~M.}\ \bibnamefont {Schloss}}, \bibinfo
  {author} {\bibfnamefont {D.~R.}\ \bibnamefont {Glenn}}, \bibinfo {author}
  {\bibfnamefont {Y.}~\bibnamefont {Song}}, \bibinfo {author} {\bibfnamefont
  {M.~D.}\ \bibnamefont {Lukin}}, \bibinfo {author} {\bibfnamefont
  {H.}~\bibnamefont {Park}},\ and\ \bibinfo {author} {\bibfnamefont {R.~L.}\
  \bibnamefont {Walsworth}},\ }\bibfield  {title} {\bibinfo {title} {{Optical
  magnetic detection of single-neuron action potentials using quantum defects
  in diamond}},\ }\href {https://doi.org/10.1073/pnas.1601513113} {\bibfield
  {journal} {\bibinfo  {journal} {Proceedings of the National Academy of
  Sciences of the United States of America}\ }\textbf {\bibinfo {volume}
  {113}},\ \bibinfo {pages} {14133} (\bibinfo {year} {2016})}\BibitemShut
  {NoStop}%
\bibitem [{\citenamefont {Hanlon}\ \emph {et~al.}(2019)\citenamefont {Hanlon},
  \citenamefont {Gautam}, \citenamefont {Wood}, \citenamefont {Reddy},
  \citenamefont {Barson}, \citenamefont {Niihori}, \citenamefont {Silalahi},
  \citenamefont {Corry}, \citenamefont {Wrachtrup}, \citenamefont {Sellars},
  \citenamefont {Daria}, \citenamefont {Maletinsky}, \citenamefont {Stuart},\
  and\ \citenamefont {Doherty}}]{hanlon2019diamond}%
  \BibitemOpen
  \bibfield  {author} {\bibinfo {author} {\bibfnamefont {L.}~\bibnamefont
  {Hanlon}}, \bibinfo {author} {\bibfnamefont {V.}~\bibnamefont {Gautam}},
  \bibinfo {author} {\bibfnamefont {J.~D.~A.}\ \bibnamefont {Wood}}, \bibinfo
  {author} {\bibfnamefont {P.}~\bibnamefont {Reddy}}, \bibinfo {author}
  {\bibfnamefont {M.~S.~J.}\ \bibnamefont {Barson}}, \bibinfo {author}
  {\bibfnamefont {M.}~\bibnamefont {Niihori}}, \bibinfo {author} {\bibfnamefont
  {A.~R.~J.}\ \bibnamefont {Silalahi}}, \bibinfo {author} {\bibfnamefont
  {B.}~\bibnamefont {Corry}}, \bibinfo {author} {\bibfnamefont
  {J.}~\bibnamefont {Wrachtrup}}, \bibinfo {author} {\bibfnamefont {M.~J.}\
  \bibnamefont {Sellars}}, \bibinfo {author} {\bibfnamefont {V.~R.}\
  \bibnamefont {Daria}}, \bibinfo {author} {\bibfnamefont {P.}~\bibnamefont
  {Maletinsky}}, \bibinfo {author} {\bibfnamefont {G.~J.}\ \bibnamefont
  {Stuart}},\ and\ \bibinfo {author} {\bibfnamefont {M.~W.}\ \bibnamefont
  {Doherty}},\ }\href@noop {} {\bibinfo {title} {{Diamond nano-pillar arrays
  for quantum microscopy of neuronal signals}}} (\bibinfo {year} {2019}),\
  \Eprint {https://arxiv.org/abs/1901.08743} {arXiv:1901.08743 [quant-ph]}
  \BibitemShut {NoStop}%
\bibitem [{\citenamefont {Schwierz}(2010)}]{Schwierz2010}%
  \BibitemOpen
  \bibfield  {author} {\bibinfo {author} {\bibfnamefont {F.}~\bibnamefont
  {Schwierz}},\ }\bibfield  {title} {\bibinfo {title} {{Graphene
  transistors}},\ }\href {https://doi.org/10.1038/nnano.2010.89} {\bibfield
  {journal} {\bibinfo  {journal} {Nature Nanotechnology}\ }\textbf {\bibinfo
  {volume} {5}},\ \bibinfo {pages} {487} (\bibinfo {year} {2010})}\BibitemShut
  {NoStop}%
\bibitem [{\citenamefont {Radisavljevic}\ \emph {et~al.}(2011)\citenamefont
  {Radisavljevic}, \citenamefont {Radenovic}, \citenamefont {Brivio},
  \citenamefont {Giacometti},\ and\ \citenamefont {Kis}}]{Radisavljevic2011}%
  \BibitemOpen
  \bibfield  {author} {\bibinfo {author} {\bibfnamefont {B.}~\bibnamefont
  {Radisavljevic}}, \bibinfo {author} {\bibfnamefont {A.}~\bibnamefont
  {Radenovic}}, \bibinfo {author} {\bibfnamefont {J.}~\bibnamefont {Brivio}},
  \bibinfo {author} {\bibfnamefont {V.}~\bibnamefont {Giacometti}},\ and\
  \bibinfo {author} {\bibfnamefont {A.}~\bibnamefont {Kis}},\ }\bibfield
  {title} {\bibinfo {title} {{Single-layer MoS2 transistors}},\ }\href
  {https://doi.org/10.1038/nnano.2010.279} {\bibfield  {journal} {\bibinfo
  {journal} {Nature Nanotechnology}\ }\textbf {\bibinfo {volume} {6}},\
  \bibinfo {pages} {147} (\bibinfo {year} {2011})}\BibitemShut {NoStop}%
\bibitem [{\citenamefont {Mak}\ and\ \citenamefont {Shan}(2016)}]{Mak2016}%
  \BibitemOpen
  \bibfield  {author} {\bibinfo {author} {\bibfnamefont {K.~F.}\ \bibnamefont
  {Mak}}\ and\ \bibinfo {author} {\bibfnamefont {J.}~\bibnamefont {Shan}},\
  }\bibfield  {title} {\bibinfo {title} {{Photonics and optoelectronics of 2D
  semiconductor transition metal dichalcogenides}},\ }\href
  {https://doi.org/10.1038/nphoton.2015.282 http://10.0.4.14/nphoton.2015.282}
  {\bibfield  {journal} {\bibinfo  {journal} {Nature Photonics}\ }\textbf
  {\bibinfo {volume} {10}},\ \bibinfo {pages} {216} (\bibinfo {year}
  {2016})}\BibitemShut {NoStop}%
\bibitem [{\citenamefont {Tao}\ \emph {et~al.}(2015)\citenamefont {Tao},
  \citenamefont {Cinquanta}, \citenamefont {Chiappe}, \citenamefont
  {Grazianetti}, \citenamefont {Fanciulli}, \citenamefont {Dubey},
  \citenamefont {Molle},\ and\ \citenamefont {Akinwande}}]{Tao2015}%
  \BibitemOpen
  \bibfield  {author} {\bibinfo {author} {\bibfnamefont {L.}~\bibnamefont
  {Tao}}, \bibinfo {author} {\bibfnamefont {E.}~\bibnamefont {Cinquanta}},
  \bibinfo {author} {\bibfnamefont {D.}~\bibnamefont {Chiappe}}, \bibinfo
  {author} {\bibfnamefont {C.}~\bibnamefont {Grazianetti}}, \bibinfo {author}
  {\bibfnamefont {M.}~\bibnamefont {Fanciulli}}, \bibinfo {author}
  {\bibfnamefont {M.}~\bibnamefont {Dubey}}, \bibinfo {author} {\bibfnamefont
  {A.}~\bibnamefont {Molle}},\ and\ \bibinfo {author} {\bibfnamefont
  {D.}~\bibnamefont {Akinwande}},\ }\bibfield  {title} {\bibinfo {title}
  {{Silicene field-effect transistors operating at room temperature}},\ }\href
  {https://doi.org/10.1038/nnano.2014.325 http://10.0.4.14/nnano.2014.325
  https://www.nature.com/articles/nnano.2014.325{\#}supplementary-information}
  {\bibfield  {journal} {\bibinfo  {journal} {Nature Nanotechnology}\ }\textbf
  {\bibinfo {volume} {10}},\ \bibinfo {pages} {227} (\bibinfo {year}
  {2015})}\BibitemShut {NoStop}%
\bibitem [{\citenamefont {Mak}\ \emph {et~al.}(2012)\citenamefont {Mak},
  \citenamefont {He}, \citenamefont {Lee}, \citenamefont {Lee}, \citenamefont
  {Hone}, \citenamefont {Heinz},\ and\ \citenamefont {Shan}}]{Mak2012}%
  \BibitemOpen
  \bibfield  {author} {\bibinfo {author} {\bibfnamefont {K.~F.}\ \bibnamefont
  {Mak}}, \bibinfo {author} {\bibfnamefont {K.}~\bibnamefont {He}}, \bibinfo
  {author} {\bibfnamefont {C.}~\bibnamefont {Lee}}, \bibinfo {author}
  {\bibfnamefont {G.~H.}\ \bibnamefont {Lee}}, \bibinfo {author} {\bibfnamefont
  {J.}~\bibnamefont {Hone}}, \bibinfo {author} {\bibfnamefont {T.~F.}\
  \bibnamefont {Heinz}},\ and\ \bibinfo {author} {\bibfnamefont
  {J.}~\bibnamefont {Shan}},\ }\bibfield  {title} {\bibinfo {title} {{Tightly
  bound trions in monolayer MoS2}},\ }\href {https://doi.org/10.1038/nmat3505
  http://10.0.4.14/nmat3505
  https://www.nature.com/articles/nmat3505{\#}supplementary-information}
  {\bibfield  {journal} {\bibinfo  {journal} {Nature Materials}\ }\textbf
  {\bibinfo {volume} {12}},\ \bibinfo {pages} {207} (\bibinfo {year}
  {2012})}\BibitemShut {NoStop}%
\bibitem [{\citenamefont {Das}\ \emph {et~al.}(2019)\citenamefont {Das},
  \citenamefont {Tang}, \citenamefont {Hong}, \citenamefont {Gon{\c{c}}alves},
  \citenamefont {McCarter}, \citenamefont {Klewe}, \citenamefont {Nguyen},
  \citenamefont {G{\'{o}}mez-Ortiz}, \citenamefont {Shafer}, \citenamefont
  {Arenholz}, \citenamefont {Stoica}, \citenamefont {Hsu}, \citenamefont
  {Wang}, \citenamefont {Ophus}, \citenamefont {Liu}, \citenamefont {Nelson},
  \citenamefont {Saremi}, \citenamefont {Prasad}, \citenamefont {Mei},
  \citenamefont {Schlom}, \citenamefont {{\'{I}}{\~{n}}iguez}, \citenamefont
  {Garc{\'{i}}a-Fern{\'{a}}ndez}, \citenamefont {Muller}, \citenamefont {Chen},
  \citenamefont {Junquera}, \citenamefont {Martin},\ and\ \citenamefont
  {Ramesh}}]{Das2019}%
  \BibitemOpen
  \bibfield  {author} {\bibinfo {author} {\bibfnamefont {S.}~\bibnamefont
  {Das}}, \bibinfo {author} {\bibfnamefont {Y.~L.}\ \bibnamefont {Tang}},
  \bibinfo {author} {\bibfnamefont {Z.}~\bibnamefont {Hong}}, \bibinfo {author}
  {\bibfnamefont {M.~A.}\ \bibnamefont {Gon{\c{c}}alves}}, \bibinfo {author}
  {\bibfnamefont {M.~R.}\ \bibnamefont {McCarter}}, \bibinfo {author}
  {\bibfnamefont {C.}~\bibnamefont {Klewe}}, \bibinfo {author} {\bibfnamefont
  {K.~X.}\ \bibnamefont {Nguyen}}, \bibinfo {author} {\bibfnamefont
  {F.}~\bibnamefont {G{\'{o}}mez-Ortiz}}, \bibinfo {author} {\bibfnamefont
  {P.}~\bibnamefont {Shafer}}, \bibinfo {author} {\bibfnamefont
  {E.}~\bibnamefont {Arenholz}}, \bibinfo {author} {\bibfnamefont {V.~A.}\
  \bibnamefont {Stoica}}, \bibinfo {author} {\bibfnamefont {S.~L.}\
  \bibnamefont {Hsu}}, \bibinfo {author} {\bibfnamefont {B.}~\bibnamefont
  {Wang}}, \bibinfo {author} {\bibfnamefont {C.}~\bibnamefont {Ophus}},
  \bibinfo {author} {\bibfnamefont {J.~F.}\ \bibnamefont {Liu}}, \bibinfo
  {author} {\bibfnamefont {C.~T.}\ \bibnamefont {Nelson}}, \bibinfo {author}
  {\bibfnamefont {S.}~\bibnamefont {Saremi}}, \bibinfo {author} {\bibfnamefont
  {B.}~\bibnamefont {Prasad}}, \bibinfo {author} {\bibfnamefont {A.~B.}\
  \bibnamefont {Mei}}, \bibinfo {author} {\bibfnamefont {D.~G.}\ \bibnamefont
  {Schlom}}, \bibinfo {author} {\bibfnamefont {J.}~\bibnamefont
  {{\'{I}}{\~{n}}iguez}}, \bibinfo {author} {\bibfnamefont {P.}~\bibnamefont
  {Garc{\'{i}}a-Fern{\'{a}}ndez}}, \bibinfo {author} {\bibfnamefont {D.~A.}\
  \bibnamefont {Muller}}, \bibinfo {author} {\bibfnamefont {L.~Q.}\
  \bibnamefont {Chen}}, \bibinfo {author} {\bibfnamefont {J.}~\bibnamefont
  {Junquera}}, \bibinfo {author} {\bibfnamefont {L.~W.}\ \bibnamefont
  {Martin}},\ and\ \bibinfo {author} {\bibfnamefont {R.}~\bibnamefont
  {Ramesh}},\ }\bibfield  {title} {\bibinfo {title} {{Observation of
  room-temperature polar skyrmions}},\ }\href
  {https://doi.org/10.1038/s41586-019-1092-8} {\bibfield  {journal} {\bibinfo
  {journal} {Nature}\ }\textbf {\bibinfo {volume} {568}},\ \bibinfo {pages}
  {368} (\bibinfo {year} {2019})}\BibitemShut {NoStop}%
\bibitem [{\citenamefont {Doherty}\ \emph {et~al.}(2013)\citenamefont
  {Doherty}, \citenamefont {Manson}, \citenamefont {Delaney}, \citenamefont
  {Jelezko}, \citenamefont {Wrachtrup},\ and\ \citenamefont
  {Hollenberg}}]{Doherty2013}%
  \BibitemOpen
  \bibfield  {author} {\bibinfo {author} {\bibfnamefont {M.~W.}\ \bibnamefont
  {Doherty}}, \bibinfo {author} {\bibfnamefont {N.~B.}\ \bibnamefont {Manson}},
  \bibinfo {author} {\bibfnamefont {P.}~\bibnamefont {Delaney}}, \bibinfo
  {author} {\bibfnamefont {F.}~\bibnamefont {Jelezko}}, \bibinfo {author}
  {\bibfnamefont {J.}~\bibnamefont {Wrachtrup}},\ and\ \bibinfo {author}
  {\bibfnamefont {L.~C.}\ \bibnamefont {Hollenberg}},\ }\bibfield  {title}
  {\bibinfo {title} {{The nitrogen-vacancy colour centre in diamond}},\ }\href
  {https://doi.org/10.1016/j.physrep.2013.02.001} {\bibfield  {journal}
  {\bibinfo  {journal} {Physics Reports}\ }\textbf {\bibinfo {volume} {528}},\
  \bibinfo {pages} {1} (\bibinfo {year} {2013})},\ \Eprint
  {https://arxiv.org/abs/1302.3288} {arXiv:1302.3288} \BibitemShut {NoStop}%
\bibitem [{\citenamefont {Dolde}\ \emph {et~al.}(2011)\citenamefont {Dolde},
  \citenamefont {Fedder}, \citenamefont {Doherty}, \citenamefont
  {N{\"{o}}bauer}, \citenamefont {Rempp}, \citenamefont {Balasubramanian},
  \citenamefont {Wolf}, \citenamefont {Reinhard}, \citenamefont {Hollenberg},
  \citenamefont {Jelezko},\ and\ \citenamefont {Wrachtrup}}]{Dolde2011a}%
  \BibitemOpen
  \bibfield  {author} {\bibinfo {author} {\bibfnamefont {F.}~\bibnamefont
  {Dolde}}, \bibinfo {author} {\bibfnamefont {H.}~\bibnamefont {Fedder}},
  \bibinfo {author} {\bibfnamefont {M.~W.}\ \bibnamefont {Doherty}}, \bibinfo
  {author} {\bibfnamefont {T.}~\bibnamefont {N{\"{o}}bauer}}, \bibinfo {author}
  {\bibfnamefont {F.}~\bibnamefont {Rempp}}, \bibinfo {author} {\bibfnamefont
  {G.}~\bibnamefont {Balasubramanian}}, \bibinfo {author} {\bibfnamefont
  {T.}~\bibnamefont {Wolf}}, \bibinfo {author} {\bibfnamefont {F.}~\bibnamefont
  {Reinhard}}, \bibinfo {author} {\bibfnamefont {L.~C.}\ \bibnamefont
  {Hollenberg}}, \bibinfo {author} {\bibfnamefont {F.}~\bibnamefont
  {Jelezko}},\ and\ \bibinfo {author} {\bibfnamefont {J.}~\bibnamefont
  {Wrachtrup}},\ }\bibfield  {title} {\bibinfo {title} {{Electric-field sensing
  using single diamond spins}},\ }\href {https://doi.org/10.1038/nphys1969}
  {\bibfield  {journal} {\bibinfo  {journal} {Nature Physics}\ }\textbf
  {\bibinfo {volume} {7}},\ \bibinfo {pages} {459} (\bibinfo {year}
  {2011})}\BibitemShut {NoStop}%
\bibitem [{\citenamefont {Dolde}\ \emph {et~al.}(2014)\citenamefont {Dolde},
  \citenamefont {Doherty}, \citenamefont {Michl}, \citenamefont {Jakobi},
  \citenamefont {Naydenov}, \citenamefont {Pezzagna}, \citenamefont {Meijer},
  \citenamefont {Neumann}, \citenamefont {Jelezko}, \citenamefont {Manson},\
  and\ \citenamefont {Wrachtrup}}]{Dolde2014a}%
  \BibitemOpen
  \bibfield  {author} {\bibinfo {author} {\bibfnamefont {F.}~\bibnamefont
  {Dolde}}, \bibinfo {author} {\bibfnamefont {M.~W.}\ \bibnamefont {Doherty}},
  \bibinfo {author} {\bibfnamefont {J.}~\bibnamefont {Michl}}, \bibinfo
  {author} {\bibfnamefont {I.}~\bibnamefont {Jakobi}}, \bibinfo {author}
  {\bibfnamefont {B.}~\bibnamefont {Naydenov}}, \bibinfo {author}
  {\bibfnamefont {S.}~\bibnamefont {Pezzagna}}, \bibinfo {author}
  {\bibfnamefont {J.}~\bibnamefont {Meijer}}, \bibinfo {author} {\bibfnamefont
  {P.}~\bibnamefont {Neumann}}, \bibinfo {author} {\bibfnamefont
  {F.}~\bibnamefont {Jelezko}}, \bibinfo {author} {\bibfnamefont {N.~B.}\
  \bibnamefont {Manson}},\ and\ \bibinfo {author} {\bibfnamefont
  {J.}~\bibnamefont {Wrachtrup}},\ }\bibfield  {title} {\bibinfo {title}
  {{Nanoscale detection of a single fundamental charge in ambient conditions
  using the NV - Center in diamond}},\ }\bibfield  {journal} {\bibinfo
  {journal} {Physical Review Letters}\ }\textbf {\bibinfo {volume} {112}},\
  \href {https://doi.org/10.1103/PhysRevLett.112.097603}
  {10.1103/PhysRevLett.112.097603} (\bibinfo {year} {2014})\BibitemShut
  {NoStop}%
\bibitem [{\citenamefont {Chen}\ \emph {et~al.}(2017)\citenamefont {Chen},
  \citenamefont {Clevenson}, \citenamefont {Johnson}, \citenamefont {Pham},
  \citenamefont {Englund}, \citenamefont {Hemmer},\ and\ \citenamefont
  {Braje}}]{Chen2017}%
  \BibitemOpen
  \bibfield  {author} {\bibinfo {author} {\bibfnamefont {E.~H.}\ \bibnamefont
  {Chen}}, \bibinfo {author} {\bibfnamefont {H.~A.}\ \bibnamefont {Clevenson}},
  \bibinfo {author} {\bibfnamefont {K.~A.}\ \bibnamefont {Johnson}}, \bibinfo
  {author} {\bibfnamefont {L.~M.}\ \bibnamefont {Pham}}, \bibinfo {author}
  {\bibfnamefont {D.~R.}\ \bibnamefont {Englund}}, \bibinfo {author}
  {\bibfnamefont {P.~R.}\ \bibnamefont {Hemmer}},\ and\ \bibinfo {author}
  {\bibfnamefont {D.~A.}\ \bibnamefont {Braje}},\ }\bibfield  {title} {\bibinfo
  {title} {{High-sensitivity spin-based electrometry with an ensemble of
  nitrogen-vacancy centers in diamond}},\ }\bibfield  {journal} {\bibinfo
  {journal} {Physical Review A}\ }\textbf {\bibinfo {volume} {95}},\ \href
  {https://doi.org/10.1103/PhysRevA.95.053417} {10.1103/PhysRevA.95.053417}
  (\bibinfo {year} {2017})\BibitemShut {NoStop}%
\bibitem [{\citenamefont {Iwasaki}\ \emph {et~al.}(2017)\citenamefont
  {Iwasaki}, \citenamefont {Naruki}, \citenamefont {Tahara}, \citenamefont
  {Makino}, \citenamefont {Kato}, \citenamefont {Ogura}, \citenamefont
  {Takeuchi}, \citenamefont {Yamasaki},\ and\ \citenamefont
  {Hatano}}]{Iwasaki2017}%
  \BibitemOpen
  \bibfield  {author} {\bibinfo {author} {\bibfnamefont {T.}~\bibnamefont
  {Iwasaki}}, \bibinfo {author} {\bibfnamefont {W.}~\bibnamefont {Naruki}},
  \bibinfo {author} {\bibfnamefont {K.}~\bibnamefont {Tahara}}, \bibinfo
  {author} {\bibfnamefont {T.}~\bibnamefont {Makino}}, \bibinfo {author}
  {\bibfnamefont {H.}~\bibnamefont {Kato}}, \bibinfo {author} {\bibfnamefont
  {M.}~\bibnamefont {Ogura}}, \bibinfo {author} {\bibfnamefont
  {D.}~\bibnamefont {Takeuchi}}, \bibinfo {author} {\bibfnamefont
  {S.}~\bibnamefont {Yamasaki}},\ and\ \bibinfo {author} {\bibfnamefont
  {M.}~\bibnamefont {Hatano}},\ }\bibfield  {title} {\bibinfo {title} {{Direct
  Nanoscale Sensing of the Internal Electric Field in Operating Semiconductor
  Devices Using Single Electron Spins}},\ }\href
  {https://doi.org/10.1021/acsnano.6b04460} {\bibfield  {journal} {\bibinfo
  {journal} {ACS Nano}\ }\textbf {\bibinfo {volume} {11}},\ \bibinfo {pages}
  {1238} (\bibinfo {year} {2017})}\BibitemShut {NoStop}%
\bibitem [{\citenamefont {Balasubramanian}\ \emph {et~al.}(2009)\citenamefont
  {Balasubramanian}, \citenamefont {Neumann}, \citenamefont {Twitchen},
  \citenamefont {Markham}, \citenamefont {Kolesov}, \citenamefont {Mizuochi},
  \citenamefont {Isoya}, \citenamefont {Achard}, \citenamefont {Beck},
  \citenamefont {Tissler}, \citenamefont {Jacques}, \citenamefont {Hemmer},
  \citenamefont {Jelezko},\ and\ \citenamefont
  {Wrachtrup}}]{Balasubramanian2009}%
  \BibitemOpen
  \bibfield  {author} {\bibinfo {author} {\bibfnamefont {G.}~\bibnamefont
  {Balasubramanian}}, \bibinfo {author} {\bibfnamefont {P.}~\bibnamefont
  {Neumann}}, \bibinfo {author} {\bibfnamefont {D.}~\bibnamefont {Twitchen}},
  \bibinfo {author} {\bibfnamefont {M.}~\bibnamefont {Markham}}, \bibinfo
  {author} {\bibfnamefont {R.}~\bibnamefont {Kolesov}}, \bibinfo {author}
  {\bibfnamefont {N.}~\bibnamefont {Mizuochi}}, \bibinfo {author}
  {\bibfnamefont {J.}~\bibnamefont {Isoya}}, \bibinfo {author} {\bibfnamefont
  {J.}~\bibnamefont {Achard}}, \bibinfo {author} {\bibfnamefont
  {J.}~\bibnamefont {Beck}}, \bibinfo {author} {\bibfnamefont {J.}~\bibnamefont
  {Tissler}}, \bibinfo {author} {\bibfnamefont {V.}~\bibnamefont {Jacques}},
  \bibinfo {author} {\bibfnamefont {P.~R.}\ \bibnamefont {Hemmer}}, \bibinfo
  {author} {\bibfnamefont {F.}~\bibnamefont {Jelezko}},\ and\ \bibinfo {author}
  {\bibfnamefont {J.}~\bibnamefont {Wrachtrup}},\ }\bibfield  {title} {\bibinfo
  {title} {{Ultralong spin coherence time in isotopically engineered
  diamond}},\ }\href {https://doi.org/10.1038/nmat2420} {\bibfield  {journal}
  {\bibinfo  {journal} {Nature Materials}\ }\textbf {\bibinfo {volume} {8}},\
  \bibinfo {pages} {383} (\bibinfo {year} {2009})}\BibitemShut {NoStop}%
\bibitem [{\citenamefont {Tetienne}\ \emph {et~al.}(2014)\citenamefont
  {Tetienne}, \citenamefont {Hingant}, \citenamefont {Kim}, \citenamefont
  {{Herrera Diez}}, \citenamefont {Adam}, \citenamefont {Garcia}, \citenamefont
  {Roch}, \citenamefont {Rohart}, \citenamefont {Thiaville}, \citenamefont
  {Ravelosona},\ and\ \citenamefont {Jacques}}]{Tetienne2014a}%
  \BibitemOpen
  \bibfield  {author} {\bibinfo {author} {\bibfnamefont {J.~P.}\ \bibnamefont
  {Tetienne}}, \bibinfo {author} {\bibfnamefont {T.}~\bibnamefont {Hingant}},
  \bibinfo {author} {\bibfnamefont {J.~V.}\ \bibnamefont {Kim}}, \bibinfo
  {author} {\bibfnamefont {L.}~\bibnamefont {{Herrera Diez}}}, \bibinfo
  {author} {\bibfnamefont {J.~P.}\ \bibnamefont {Adam}}, \bibinfo {author}
  {\bibfnamefont {K.}~\bibnamefont {Garcia}}, \bibinfo {author} {\bibfnamefont
  {J.~F.}\ \bibnamefont {Roch}}, \bibinfo {author} {\bibfnamefont
  {S.}~\bibnamefont {Rohart}}, \bibinfo {author} {\bibfnamefont
  {A.}~\bibnamefont {Thiaville}}, \bibinfo {author} {\bibfnamefont
  {D.}~\bibnamefont {Ravelosona}},\ and\ \bibinfo {author} {\bibfnamefont
  {V.}~\bibnamefont {Jacques}},\ }\bibfield  {title} {\bibinfo {title}
  {{Nanoscale imaging and control of domain-wall hopping with a
  nitrogen-vacancy center microscope}},\ }\href
  {https://doi.org/10.1126/science.1250113} {\bibfield  {journal} {\bibinfo
  {journal} {Science}\ }\textbf {\bibinfo {volume} {344}},\ \bibinfo {pages}
  {1366} (\bibinfo {year} {2014})}\BibitemShut {NoStop}%
\bibitem [{\citenamefont {Thiel}\ \emph {et~al.}(2016)\citenamefont {Thiel},
  \citenamefont {Rohner}, \citenamefont {Ganzhorn}, \citenamefont {Appel},
  \citenamefont {Neu}, \citenamefont {M{\"{u}}ller}, \citenamefont {Kleiner},
  \citenamefont {Koelle},\ and\ \citenamefont {Maletinsky}}]{Thiel2016}%
  \BibitemOpen
  \bibfield  {author} {\bibinfo {author} {\bibfnamefont {L.}~\bibnamefont
  {Thiel}}, \bibinfo {author} {\bibfnamefont {D.}~\bibnamefont {Rohner}},
  \bibinfo {author} {\bibfnamefont {M.}~\bibnamefont {Ganzhorn}}, \bibinfo
  {author} {\bibfnamefont {P.}~\bibnamefont {Appel}}, \bibinfo {author}
  {\bibfnamefont {E.}~\bibnamefont {Neu}}, \bibinfo {author} {\bibfnamefont
  {B.}~\bibnamefont {M{\"{u}}ller}}, \bibinfo {author} {\bibfnamefont
  {R.}~\bibnamefont {Kleiner}}, \bibinfo {author} {\bibfnamefont
  {D.}~\bibnamefont {Koelle}},\ and\ \bibinfo {author} {\bibfnamefont
  {P.}~\bibnamefont {Maletinsky}},\ }\bibfield  {title} {\bibinfo {title}
  {{Quantitative nanoscale vortex imaging using a cryogenic quantum
  magnetometer}},\ }\href {https://doi.org/10.1038/nnano.2016.63} {\bibfield
  {journal} {\bibinfo  {journal} {Nature Nanotechnology}\ }\textbf {\bibinfo
  {volume} {11}},\ \bibinfo {pages} {677} (\bibinfo {year} {2016})},\ \Eprint
  {https://arxiv.org/abs/1511.02873} {arXiv:1511.02873} \BibitemShut {NoStop}%
\bibitem [{\citenamefont {H{\"{a}}berle}\ \emph {et~al.}(2015)\citenamefont
  {H{\"{a}}berle}, \citenamefont {Schmid-Lorch}, \citenamefont {Reinhard},\
  and\ \citenamefont {Wrachtrup}}]{Haberle2015}%
  \BibitemOpen
  \bibfield  {author} {\bibinfo {author} {\bibfnamefont {T.}~\bibnamefont
  {H{\"{a}}berle}}, \bibinfo {author} {\bibfnamefont {D.}~\bibnamefont
  {Schmid-Lorch}}, \bibinfo {author} {\bibfnamefont {F.}~\bibnamefont
  {Reinhard}},\ and\ \bibinfo {author} {\bibfnamefont {J.}~\bibnamefont
  {Wrachtrup}},\ }\bibfield  {title} {\bibinfo {title} {{Nanoscale nuclear
  magnetic imaging with chemical contrast}},\ }\href
  {https://doi.org/10.1038/nnano.2014.299} {\bibfield  {journal} {\bibinfo
  {journal} {Nature Nanotechnology}\ }\textbf {\bibinfo {volume} {10}},\
  \bibinfo {pages} {125} (\bibinfo {year} {2015})}\BibitemShut {NoStop}%
\bibitem [{\citenamefont {Arai}\ \emph {et~al.}(2015)\citenamefont {Arai},
  \citenamefont {Belthangady}, \citenamefont {Zhang}, \citenamefont {Bar-Gill},
  \citenamefont {DeVience}, \citenamefont {Cappellaro}, \citenamefont
  {Yacoby},\ and\ \citenamefont {Walsworth}}]{Arai2015}%
  \BibitemOpen
  \bibfield  {author} {\bibinfo {author} {\bibfnamefont {K.}~\bibnamefont
  {Arai}}, \bibinfo {author} {\bibfnamefont {C.}~\bibnamefont {Belthangady}},
  \bibinfo {author} {\bibfnamefont {H.}~\bibnamefont {Zhang}}, \bibinfo
  {author} {\bibfnamefont {N.}~\bibnamefont {Bar-Gill}}, \bibinfo {author}
  {\bibfnamefont {S.~J.}\ \bibnamefont {DeVience}}, \bibinfo {author}
  {\bibfnamefont {P.}~\bibnamefont {Cappellaro}}, \bibinfo {author}
  {\bibfnamefont {A.}~\bibnamefont {Yacoby}},\ and\ \bibinfo {author}
  {\bibfnamefont {R.~L.}\ \bibnamefont {Walsworth}},\ }\bibfield  {title}
  {\bibinfo {title} {{Fourier magnetic imaging with nanoscale resolution and
  compressed sensing speed-up using electronic spins in diamond}},\ }\href
  {https://doi.org/10.1038/nnano.2015.171} {\bibfield  {journal} {\bibinfo
  {journal} {Nature Nanotechnology}\ }\textbf {\bibinfo {volume} {10}},\
  \bibinfo {pages} {859} (\bibinfo {year} {2015})}\BibitemShut {NoStop}%
\bibitem [{\citenamefont {Kucsko}\ \emph {et~al.}(2013)\citenamefont {Kucsko},
  \citenamefont {Maurer}, \citenamefont {Yao}, \citenamefont {Kubo},
  \citenamefont {Noh}, \citenamefont {Lo}, \citenamefont {Park},\ and\
  \citenamefont {Lukin}}]{Kucsko2013}%
  \BibitemOpen
  \bibfield  {author} {\bibinfo {author} {\bibfnamefont {G.}~\bibnamefont
  {Kucsko}}, \bibinfo {author} {\bibfnamefont {P.~C.}\ \bibnamefont {Maurer}},
  \bibinfo {author} {\bibfnamefont {N.~Y.}\ \bibnamefont {Yao}}, \bibinfo
  {author} {\bibfnamefont {M.}~\bibnamefont {Kubo}}, \bibinfo {author}
  {\bibfnamefont {H.~J.}\ \bibnamefont {Noh}}, \bibinfo {author} {\bibfnamefont
  {P.~K.}\ \bibnamefont {Lo}}, \bibinfo {author} {\bibfnamefont
  {H.}~\bibnamefont {Park}},\ and\ \bibinfo {author} {\bibfnamefont {M.~D.}\
  \bibnamefont {Lukin}},\ }\bibfield  {title} {\bibinfo {title}
  {{Nanometre-scale thermometry in a living cell}},\ }\href
  {https://doi.org/10.1038/nature12373} {\bibfield  {journal} {\bibinfo
  {journal} {Nature}\ }\textbf {\bibinfo {volume} {500}},\ \bibinfo {pages}
  {54} (\bibinfo {year} {2013})},\ \Eprint {https://arxiv.org/abs/1304.1068}
  {arXiv:1304.1068} \BibitemShut {NoStop}%
\bibitem [{\citenamefont {Neumann}\ \emph {et~al.}(2013)\citenamefont
  {Neumann}, \citenamefont {Jakobi}, \citenamefont {Dolde}, \citenamefont
  {Burk}, \citenamefont {Reuter}, \citenamefont {Waldherr}, \citenamefont
  {Honert}, \citenamefont {Wolf}, \citenamefont {Brunner}, \citenamefont
  {Shim}, \citenamefont {Suter}, \citenamefont {Sumiya}, \citenamefont
  {Isoya},\ and\ \citenamefont {Wrachtrup}}]{Neumann2013}%
  \BibitemOpen
  \bibfield  {author} {\bibinfo {author} {\bibfnamefont {P.}~\bibnamefont
  {Neumann}}, \bibinfo {author} {\bibfnamefont {I.}~\bibnamefont {Jakobi}},
  \bibinfo {author} {\bibfnamefont {F.}~\bibnamefont {Dolde}}, \bibinfo
  {author} {\bibfnamefont {C.}~\bibnamefont {Burk}}, \bibinfo {author}
  {\bibfnamefont {R.}~\bibnamefont {Reuter}}, \bibinfo {author} {\bibfnamefont
  {G.}~\bibnamefont {Waldherr}}, \bibinfo {author} {\bibfnamefont
  {J.}~\bibnamefont {Honert}}, \bibinfo {author} {\bibfnamefont
  {T.}~\bibnamefont {Wolf}}, \bibinfo {author} {\bibfnamefont {A.}~\bibnamefont
  {Brunner}}, \bibinfo {author} {\bibfnamefont {J.~H.}\ \bibnamefont {Shim}},
  \bibinfo {author} {\bibfnamefont {D.}~\bibnamefont {Suter}}, \bibinfo
  {author} {\bibfnamefont {H.}~\bibnamefont {Sumiya}}, \bibinfo {author}
  {\bibfnamefont {J.}~\bibnamefont {Isoya}},\ and\ \bibinfo {author}
  {\bibfnamefont {J.}~\bibnamefont {Wrachtrup}},\ }\bibfield  {title} {\bibinfo
  {title} {{High-precision nanoscale temperature sensing using single defects
  in diamond}},\ }\href {https://doi.org/10.1021/nl401216y} {\bibfield
  {journal} {\bibinfo  {journal} {Nano Letters}\ }\textbf {\bibinfo {volume}
  {13}},\ \bibinfo {pages} {2738} (\bibinfo {year} {2013})},\ \Eprint
  {https://arxiv.org/abs/1304.0688} {arXiv:1304.0688} \BibitemShut {NoStop}%
\bibitem [{\citenamefont {Toyli}\ \emph {et~al.}(2013)\citenamefont {Toyli},
  \citenamefont {{De Las Casas}}, \citenamefont {Christle}, \citenamefont
  {Dobrovitski},\ and\ \citenamefont {Awschalom}}]{Toyli2013}%
  \BibitemOpen
  \bibfield  {author} {\bibinfo {author} {\bibfnamefont {D.~M.}\ \bibnamefont
  {Toyli}}, \bibinfo {author} {\bibfnamefont {C.~F.}\ \bibnamefont {{De Las
  Casas}}}, \bibinfo {author} {\bibfnamefont {D.~J.}\ \bibnamefont {Christle}},
  \bibinfo {author} {\bibfnamefont {V.~V.}\ \bibnamefont {Dobrovitski}},\ and\
  \bibinfo {author} {\bibfnamefont {D.~D.}\ \bibnamefont {Awschalom}},\
  }\bibfield  {title} {\bibinfo {title} {{Fluorescence thermometry enhanced by
  the quantum coherence of single spins in diamond}},\ }\href
  {https://doi.org/10.1073/pnas.1306825110} {\bibfield  {journal} {\bibinfo
  {journal} {Proceedings of the National Academy of Sciences of the United
  States of America}\ }\textbf {\bibinfo {volume} {110}},\ \bibinfo {pages}
  {8417} (\bibinfo {year} {2013})}\BibitemShut {NoStop}%
\bibitem [{\citenamefont {Waldherr}\ \emph {et~al.}(2014)\citenamefont
  {Waldherr}, \citenamefont {Wang}, \citenamefont {Zaiser}, \citenamefont
  {Jamali}, \citenamefont {Schulte-Herbr{\"{u}}ggen}, \citenamefont {Abe},
  \citenamefont {Ohshima}, \citenamefont {Isoya}, \citenamefont {Du},
  \citenamefont {Neumann},\ and\ \citenamefont {Wrachtrup}}]{Waldherr2014b}%
  \BibitemOpen
  \bibfield  {author} {\bibinfo {author} {\bibfnamefont {G.}~\bibnamefont
  {Waldherr}}, \bibinfo {author} {\bibfnamefont {Y.}~\bibnamefont {Wang}},
  \bibinfo {author} {\bibfnamefont {S.}~\bibnamefont {Zaiser}}, \bibinfo
  {author} {\bibfnamefont {M.}~\bibnamefont {Jamali}}, \bibinfo {author}
  {\bibfnamefont {T.}~\bibnamefont {Schulte-Herbr{\"{u}}ggen}}, \bibinfo
  {author} {\bibfnamefont {H.}~\bibnamefont {Abe}}, \bibinfo {author}
  {\bibfnamefont {T.}~\bibnamefont {Ohshima}}, \bibinfo {author} {\bibfnamefont
  {J.}~\bibnamefont {Isoya}}, \bibinfo {author} {\bibfnamefont {J.~F.}\
  \bibnamefont {Du}}, \bibinfo {author} {\bibfnamefont {P.}~\bibnamefont
  {Neumann}},\ and\ \bibinfo {author} {\bibfnamefont {J.}~\bibnamefont
  {Wrachtrup}},\ }\bibfield  {title} {\bibinfo {title} {{Quantum error
  correction in a solid-state hybrid spin register}},\ }\href
  {https://doi.org/10.1038/nature12919} {\bibfield  {journal} {\bibinfo
  {journal} {Nature}\ }\textbf {\bibinfo {volume} {506}},\ \bibinfo {pages}
  {204} (\bibinfo {year} {2014})},\ \Eprint {https://arxiv.org/abs/1309.6424}
  {arXiv:1309.6424} \BibitemShut {NoStop}%
\bibitem [{\citenamefont {Cai}\ \emph {et~al.}(2013)\citenamefont {Cai},
  \citenamefont {Retzker}, \citenamefont {Jelezko},\ and\ \citenamefont
  {Plenio}}]{Cai2013}%
  \BibitemOpen
  \bibfield  {author} {\bibinfo {author} {\bibfnamefont {J.}~\bibnamefont
  {Cai}}, \bibinfo {author} {\bibfnamefont {A.}~\bibnamefont {Retzker}},
  \bibinfo {author} {\bibfnamefont {F.}~\bibnamefont {Jelezko}},\ and\ \bibinfo
  {author} {\bibfnamefont {M.~B.}\ \bibnamefont {Plenio}},\ }\bibfield  {title}
  {\bibinfo {title} {{A large-scale quantum simulator on a diamond surface at
  room temperature}},\ }\href {https://doi.org/10.1038/nphys2519} {\bibfield
  {journal} {\bibinfo  {journal} {Nature Physics}\ }\textbf {\bibinfo {volume}
  {9}},\ \bibinfo {pages} {168} (\bibinfo {year} {2013})},\ \Eprint
  {https://arxiv.org/abs/1208.2874} {arXiv:1208.2874} \BibitemShut {NoStop}%
\bibitem [{\citenamefont {Oberg}\ \emph {et~al.}(2019)\citenamefont {Oberg},
  \citenamefont {Huang}, \citenamefont {Reddy}, \citenamefont {Alkauskas},
  \citenamefont {Greentree}, \citenamefont {Cole}, \citenamefont {Manson},
  \citenamefont {Meriles},\ and\ \citenamefont {Doherty}}]{ObergLachlanM2019}%
  \BibitemOpen
  \bibfield  {author} {\bibinfo {author} {\bibfnamefont {L.~M.}\ \bibnamefont
  {Oberg}}, \bibinfo {author} {\bibfnamefont {E.}~\bibnamefont {Huang}},
  \bibinfo {author} {\bibfnamefont {P.~M.}\ \bibnamefont {Reddy}}, \bibinfo
  {author} {\bibfnamefont {A.}~\bibnamefont {Alkauskas}}, \bibinfo {author}
  {\bibfnamefont {A.~D.}\ \bibnamefont {Greentree}}, \bibinfo {author}
  {\bibfnamefont {J.~H.}\ \bibnamefont {Cole}}, \bibinfo {author}
  {\bibfnamefont {N.~B.}\ \bibnamefont {Manson}}, \bibinfo {author}
  {\bibfnamefont {C.~A.}\ \bibnamefont {Meriles}},\ and\ \bibinfo {author}
  {\bibfnamefont {M.~W.}\ \bibnamefont {Doherty}},\ }\bibfield  {title}
  {\bibinfo {title} {{Spin coherent quantum transport of electrons between
  defects in diamond}},\ }\href {https://doi.org/10.1515/nanoph-2019-0144}
  {\bibfield  {journal} {\bibinfo  {journal} {Nanophotonics}\ }\textbf
  {\bibinfo {volume} {8}},\ \bibinfo {pages} {1975} (\bibinfo {year}
  {2019})}\BibitemShut {NoStop}%
\bibitem [{\citenamefont {Broadway}\ \emph {et~al.}(2018)\citenamefont
  {Broadway}, \citenamefont {Dontschuk}, \citenamefont {Tsai}, \citenamefont
  {Lillie}, \citenamefont {Lew}, \citenamefont {McCallum}, \citenamefont
  {Johnson}, \citenamefont {Doherty}, \citenamefont {Stacey}, \citenamefont
  {Hollenberg},\ and\ \citenamefont {Tetienne}}]{Broadway2018}%
  \BibitemOpen
  \bibfield  {author} {\bibinfo {author} {\bibfnamefont {D.~A.}\ \bibnamefont
  {Broadway}}, \bibinfo {author} {\bibfnamefont {N.}~\bibnamefont {Dontschuk}},
  \bibinfo {author} {\bibfnamefont {A.}~\bibnamefont {Tsai}}, \bibinfo {author}
  {\bibfnamefont {S.~E.}\ \bibnamefont {Lillie}}, \bibinfo {author}
  {\bibfnamefont {C.~T.-K.}\ \bibnamefont {Lew}}, \bibinfo {author}
  {\bibfnamefont {J.~C.}\ \bibnamefont {McCallum}}, \bibinfo {author}
  {\bibfnamefont {B.~C.}\ \bibnamefont {Johnson}}, \bibinfo {author}
  {\bibfnamefont {M.~W.}\ \bibnamefont {Doherty}}, \bibinfo {author}
  {\bibfnamefont {A.}~\bibnamefont {Stacey}}, \bibinfo {author} {\bibfnamefont
  {L.~C.~L.}\ \bibnamefont {Hollenberg}},\ and\ \bibinfo {author}
  {\bibfnamefont {J.-P.}\ \bibnamefont {Tetienne}},\ }\bibfield  {title}
  {\bibinfo {title} {{Spatial mapping of band bending in semiconductor devices
  using in situ quantum sensors}},\ }\href
  {https://doi.org/10.1038/s41928-018-0130-0} {\bibfield  {journal} {\bibinfo
  {journal} {Nature Electronics}\ }\textbf {\bibinfo {volume} {1}},\ \bibinfo
  {pages} {502} (\bibinfo {year} {2018})}\BibitemShut {NoStop}%
\bibitem [{\citenamefont {Stacey}\ \emph {et~al.}(2019)\citenamefont {Stacey},
  \citenamefont {Dontschuk}, \citenamefont {Chou}, \citenamefont {Broadway},
  \citenamefont {Schenk}, \citenamefont {Sear}, \citenamefont {Tetienne},
  \citenamefont {Hoffman}, \citenamefont {Prawer}, \citenamefont {Pakes},
  \citenamefont {Tadich}, \citenamefont {de~Leon}, \citenamefont {Gali},\ and\
  \citenamefont {Hollenberg}}]{Stacey2019}%
  \BibitemOpen
  \bibfield  {author} {\bibinfo {author} {\bibfnamefont {A.}~\bibnamefont
  {Stacey}}, \bibinfo {author} {\bibfnamefont {N.}~\bibnamefont {Dontschuk}},
  \bibinfo {author} {\bibfnamefont {J.~P.}\ \bibnamefont {Chou}}, \bibinfo
  {author} {\bibfnamefont {D.~A.}\ \bibnamefont {Broadway}}, \bibinfo {author}
  {\bibfnamefont {A.~K.}\ \bibnamefont {Schenk}}, \bibinfo {author}
  {\bibfnamefont {M.~J.}\ \bibnamefont {Sear}}, \bibinfo {author}
  {\bibfnamefont {J.~P.}\ \bibnamefont {Tetienne}}, \bibinfo {author}
  {\bibfnamefont {A.}~\bibnamefont {Hoffman}}, \bibinfo {author} {\bibfnamefont
  {S.}~\bibnamefont {Prawer}}, \bibinfo {author} {\bibfnamefont {C.~I.}\
  \bibnamefont {Pakes}}, \bibinfo {author} {\bibfnamefont {A.}~\bibnamefont
  {Tadich}}, \bibinfo {author} {\bibfnamefont {N.~P.}\ \bibnamefont {de~Leon}},
  \bibinfo {author} {\bibfnamefont {A.}~\bibnamefont {Gali}},\ and\ \bibinfo
  {author} {\bibfnamefont {L.~C.}\ \bibnamefont {Hollenberg}},\ }\bibfield
  {title} {\bibinfo {title} {{Evidence for Primal sp 2 Defects at the Diamond
  Surface: Candidates for Electron Trapping and Noise Sources}},\ }\bibfield
  {journal} {\bibinfo  {journal} {Advanced Materials Interfaces}\ }\textbf
  {\bibinfo {volume} {6}},\ \href {https://doi.org/10.1002/admi.201801449}
  {10.1002/admi.201801449} (\bibinfo {year} {2019})\BibitemShut {NoStop}%
\bibitem [{\citenamefont {Mertens}\ \emph {et~al.}(2016)\citenamefont
  {Mertens}, \citenamefont {Mohr}, \citenamefont {Br{\"{u}}hne}, \citenamefont
  {Fecht}, \citenamefont {{\L}ojkowski}, \citenamefont
  {{\'{S}}wi{\c{e}}szkowski},\ and\ \citenamefont
  {{\L}ojkowski}}]{Mertens2016}%
  \BibitemOpen
  \bibfield  {author} {\bibinfo {author} {\bibfnamefont {M.}~\bibnamefont
  {Mertens}}, \bibinfo {author} {\bibfnamefont {M.}~\bibnamefont {Mohr}},
  \bibinfo {author} {\bibfnamefont {K.}~\bibnamefont {Br{\"{u}}hne}}, \bibinfo
  {author} {\bibfnamefont {H.~J.}\ \bibnamefont {Fecht}}, \bibinfo {author}
  {\bibfnamefont {M.}~\bibnamefont {{\L}ojkowski}}, \bibinfo {author}
  {\bibfnamefont {W.}~\bibnamefont {{\'{S}}wi{\c{e}}szkowski}},\ and\ \bibinfo
  {author} {\bibfnamefont {W.}~\bibnamefont {{\L}ojkowski}},\ }\bibfield
  {title} {\bibinfo {title} {{Patterned hydrophobic and hydrophilic surfaces of
  ultra-smooth nanocrystalline diamond layers}},\ }\href
  {https://doi.org/10.1016/j.apsusc.2016.08.130} {\bibfield  {journal}
  {\bibinfo  {journal} {Applied Surface Science}\ }\textbf {\bibinfo {volume}
  {390}},\ \bibinfo {pages} {526} (\bibinfo {year} {2016})}\BibitemShut
  {NoStop}%
\bibitem [{\citenamefont {Liao}\ \emph {et~al.}(2008)\citenamefont {Liao},
  \citenamefont {Koide}, \citenamefont {Alvarez}, \citenamefont {Imura},\ and\
  \citenamefont {Kleider}}]{Liao2008}%
  \BibitemOpen
  \bibfield  {author} {\bibinfo {author} {\bibfnamefont {M.}~\bibnamefont
  {Liao}}, \bibinfo {author} {\bibfnamefont {Y.}~\bibnamefont {Koide}},
  \bibinfo {author} {\bibfnamefont {J.}~\bibnamefont {Alvarez}}, \bibinfo
  {author} {\bibfnamefont {M.}~\bibnamefont {Imura}},\ and\ \bibinfo {author}
  {\bibfnamefont {J.~P.}\ \bibnamefont {Kleider}},\ }\bibfield  {title}
  {\bibinfo {title} {{Persistent positive and transient absolute negative
  photoconductivity observed in diamond photodetectors}},\ }\bibfield
  {journal} {\bibinfo  {journal} {Physical Review B - Condensed Matter and
  Materials Physics}\ }\textbf {\bibinfo {volume} {78}},\ \href
  {https://doi.org/10.1103/PhysRevB.78.045112} {10.1103/PhysRevB.78.045112}
  (\bibinfo {year} {2008})\BibitemShut {NoStop}%
\bibitem [{\citenamefont {Hauf}\ \emph {et~al.}(2011)\citenamefont {Hauf},
  \citenamefont {Grotz}, \citenamefont {Naydenov}, \citenamefont {Dankerl},
  \citenamefont {Pezzagna}, \citenamefont {Meijer}, \citenamefont {Jelezko},
  \citenamefont {Wrachtrup}, \citenamefont {Stutzmann}, \citenamefont
  {Reinhard},\ and\ \citenamefont {Garrido}}]{Hauf2011}%
  \BibitemOpen
  \bibfield  {author} {\bibinfo {author} {\bibfnamefont {M.~V.}\ \bibnamefont
  {Hauf}}, \bibinfo {author} {\bibfnamefont {B.}~\bibnamefont {Grotz}},
  \bibinfo {author} {\bibfnamefont {B.}~\bibnamefont {Naydenov}}, \bibinfo
  {author} {\bibfnamefont {M.}~\bibnamefont {Dankerl}}, \bibinfo {author}
  {\bibfnamefont {S.}~\bibnamefont {Pezzagna}}, \bibinfo {author}
  {\bibfnamefont {J.}~\bibnamefont {Meijer}}, \bibinfo {author} {\bibfnamefont
  {F.}~\bibnamefont {Jelezko}}, \bibinfo {author} {\bibfnamefont
  {J.}~\bibnamefont {Wrachtrup}}, \bibinfo {author} {\bibfnamefont
  {M.}~\bibnamefont {Stutzmann}}, \bibinfo {author} {\bibfnamefont
  {F.}~\bibnamefont {Reinhard}},\ and\ \bibinfo {author} {\bibfnamefont
  {J.~A.}\ \bibnamefont {Garrido}},\ }\bibfield  {title} {\bibinfo {title}
  {{Chemical control of the charge state of nitrogen-vacancy centers in
  diamond}},\ }\bibfield  {journal} {\bibinfo  {journal} {Physical Review B -
  Condensed Matter and Materials Physics}\ }\textbf {\bibinfo {volume} {83}},\
  \href {https://doi.org/10.1103/PhysRevB.83.081304}
  {10.1103/PhysRevB.83.081304} (\bibinfo {year} {2011})\BibitemShut {NoStop}%
\bibitem [{\citenamefont {Cui}\ and\ \citenamefont {Hu}(2013)}]{Cui2013}%
  \BibitemOpen
  \bibfield  {author} {\bibinfo {author} {\bibfnamefont {S.}~\bibnamefont
  {Cui}}\ and\ \bibinfo {author} {\bibfnamefont {E.~L.}\ \bibnamefont {Hu}},\
  }\bibfield  {title} {\bibinfo {title} {{Increased negatively charged
  nitrogen-vacancy centers in fluorinated diamond}},\ }\bibfield  {journal}
  {\bibinfo  {journal} {Applied Physics Letters}\ }\textbf {\bibinfo {volume}
  {103}},\ \href {https://doi.org/10.1063/1.4817651} {10.1063/1.4817651}
  (\bibinfo {year} {2013})\BibitemShut {NoStop}%
\bibitem [{\citenamefont {Ohno}\ \emph {et~al.}(2012)\citenamefont {Ohno},
  \citenamefont {{Joseph Heremans}}, \citenamefont {Bassett}, \citenamefont
  {Myers}, \citenamefont {Toyli}, \citenamefont {{Bleszynski Jayich}},
  \citenamefont {Palmstr{\o}m},\ and\ \citenamefont {Awschalom}}]{Ohno2012a}%
  \BibitemOpen
  \bibfield  {author} {\bibinfo {author} {\bibfnamefont {K.}~\bibnamefont
  {Ohno}}, \bibinfo {author} {\bibfnamefont {F.}~\bibnamefont {{Joseph
  Heremans}}}, \bibinfo {author} {\bibfnamefont {L.~C.}\ \bibnamefont
  {Bassett}}, \bibinfo {author} {\bibfnamefont {B.~A.}\ \bibnamefont {Myers}},
  \bibinfo {author} {\bibfnamefont {D.~M.}\ \bibnamefont {Toyli}}, \bibinfo
  {author} {\bibfnamefont {A.~C.}\ \bibnamefont {{Bleszynski Jayich}}},
  \bibinfo {author} {\bibfnamefont {C.~J.}\ \bibnamefont {Palmstr{\o}m}},\ and\
  \bibinfo {author} {\bibfnamefont {D.~D.}\ \bibnamefont {Awschalom}},\
  }\bibfield  {title} {\bibinfo {title} {{Engineering shallow spins in diamond
  with nitrogen delta-doping}},\ }\bibfield  {journal} {\bibinfo  {journal}
  {Applied Physics Letters}\ }\textbf {\bibinfo {volume} {101}},\ \href
  {https://doi.org/10.1063/1.4748280} {10.1063/1.4748280} (\bibinfo {year}
  {2012})\BibitemShut {NoStop}%
\bibitem [{\citenamefont {Mayrhofer}\ \emph {et~al.}(2016)\citenamefont
  {Mayrhofer}, \citenamefont {Moras}, \citenamefont {Mulakaluri}, \citenamefont
  {Rajagopalan}, \citenamefont {Stevens},\ and\ \citenamefont
  {Moseler}}]{Mayrhofer2016}%
  \BibitemOpen
  \bibfield  {author} {\bibinfo {author} {\bibfnamefont {L.}~\bibnamefont
  {Mayrhofer}}, \bibinfo {author} {\bibfnamefont {G.}~\bibnamefont {Moras}},
  \bibinfo {author} {\bibfnamefont {N.}~\bibnamefont {Mulakaluri}}, \bibinfo
  {author} {\bibfnamefont {S.}~\bibnamefont {Rajagopalan}}, \bibinfo {author}
  {\bibfnamefont {P.~A.}\ \bibnamefont {Stevens}},\ and\ \bibinfo {author}
  {\bibfnamefont {M.}~\bibnamefont {Moseler}},\ }\bibfield  {title} {\bibinfo
  {title} {{Fluorine-Terminated Diamond Surfaces as Dense Dipole Lattices: The
  Electrostatic Origin of Polar Hydrophobicity}},\ }\href
  {https://doi.org/10.1021/jacs.5b04073} {\bibfield  {journal} {\bibinfo
  {journal} {Journal of the American Chemical Society}\ }\textbf {\bibinfo
  {volume} {138}},\ \bibinfo {pages} {4018} (\bibinfo {year}
  {2016})}\BibitemShut {NoStop}%
\bibitem [{\citenamefont {Kissa}(2001)}]{Kissa2011}%
  \BibitemOpen
  \bibfield  {author} {\bibinfo {author} {\bibfnamefont {E.}~\bibnamefont
  {Kissa}},\ }\href@noop {} {\emph {\bibinfo {title} {{Fluorinated Surfactants
  and Repellents}}}},\ \bibinfo {edition} {2nd}\ ed.,\ Vol.~\bibinfo {volume}
  {97}\ (\bibinfo  {publisher} {CRC Press},\ \bibinfo {year}
  {2001})\BibitemShut {NoStop}%
\bibitem [{\citenamefont {Rietwyk}\ \emph {et~al.}(2013)\citenamefont
  {Rietwyk}, \citenamefont {Wong}, \citenamefont {Cao}, \citenamefont
  {Odonnell}, \citenamefont {Ley}, \citenamefont {Wee},\ and\ \citenamefont
  {Pakes}}]{Rietwyk2013}%
  \BibitemOpen
  \bibfield  {author} {\bibinfo {author} {\bibfnamefont {K.~J.}\ \bibnamefont
  {Rietwyk}}, \bibinfo {author} {\bibfnamefont {S.~L.}\ \bibnamefont {Wong}},
  \bibinfo {author} {\bibfnamefont {L.}~\bibnamefont {Cao}}, \bibinfo {author}
  {\bibfnamefont {K.~M.}\ \bibnamefont {Odonnell}}, \bibinfo {author}
  {\bibfnamefont {L.}~\bibnamefont {Ley}}, \bibinfo {author} {\bibfnamefont
  {A.~T.}\ \bibnamefont {Wee}},\ and\ \bibinfo {author} {\bibfnamefont {C.~I.}\
  \bibnamefont {Pakes}},\ }\bibfield  {title} {\bibinfo {title} {{Work function
  and electron affinity of the fluorine-terminated (100) diamond surface}},\
  }\bibfield  {journal} {\bibinfo  {journal} {Applied Physics Letters}\
  }\textbf {\bibinfo {volume} {102}},\ \href
  {https://doi.org/10.1063/1.4793999} {10.1063/1.4793999} (\bibinfo {year}
  {2013})\BibitemShut {NoStop}%
\bibitem [{\citenamefont {Ridley}(2013)}]{FRS2013}%
  \BibitemOpen
  \bibfield  {author} {\bibinfo {author} {\bibfnamefont {B.~K.}\ \bibnamefont
  {Ridley}},\ }\bibfield  {title} {\bibinfo {title} {{Quantum Processes in
  Semiconductors}},\ }\href {https://doi.org/10.1017/CBO9781107415324.004}
  {\bibfield  {journal} {\bibinfo  {journal} {Journal of Chemical Information
  and Modeling}\ }\textbf {\bibinfo {volume} {53}},\ \bibinfo {pages} {1689}
  (\bibinfo {year} {2013})},\ \Eprint {https://arxiv.org/abs/arXiv:1011.1669v3}
  {arXiv:arXiv:1011.1669v3} \BibitemShut {NoStop}%
\bibitem [{\citenamefont {Widmann}\ \emph {et~al.}(2014)\citenamefont
  {Widmann}, \citenamefont {Giese}, \citenamefont {Wolfer}, \citenamefont
  {Kono},\ and\ \citenamefont {Nebel}}]{Widmann2014}%
  \BibitemOpen
  \bibfield  {author} {\bibinfo {author} {\bibfnamefont {C.~J.}\ \bibnamefont
  {Widmann}}, \bibinfo {author} {\bibfnamefont {C.}~\bibnamefont {Giese}},
  \bibinfo {author} {\bibfnamefont {M.}~\bibnamefont {Wolfer}}, \bibinfo
  {author} {\bibfnamefont {S.}~\bibnamefont {Kono}},\ and\ \bibinfo {author}
  {\bibfnamefont {C.~E.}\ \bibnamefont {Nebel}},\ }\bibfield  {title} {\bibinfo
  {title} {{F-and Cl-terminations of (100) oriented single crystalline
  diamond}},\ }\href {https://doi.org/10.1002/pssa.201431188} {\bibfield
  {journal} {\bibinfo  {journal} {Physica Status Solidi (A) Applications and
  Materials Science}\ }\textbf {\bibinfo {volume} {211}},\ \bibinfo {pages}
  {2328} (\bibinfo {year} {2014})}\BibitemShut {NoStop}%
\bibitem [{\citenamefont {Brunauer}\ \emph {et~al.}(1938)\citenamefont
  {Brunauer}, \citenamefont {Emmett},\ and\ \citenamefont
  {Teller}}]{Brunauer1938}%
  \BibitemOpen
  \bibfield  {author} {\bibinfo {author} {\bibfnamefont {S.}~\bibnamefont
  {Brunauer}}, \bibinfo {author} {\bibfnamefont {P.~H.}\ \bibnamefont
  {Emmett}},\ and\ \bibinfo {author} {\bibfnamefont {E.}~\bibnamefont
  {Teller}},\ }\bibfield  {title} {\bibinfo {title} {{Adsorption of Gases in
  Multimolecular Layers}},\ }\href {https://doi.org/10.1021/ja01269a023}
  {\bibfield  {journal} {\bibinfo  {journal} {Journal of the American Chemical
  Society}\ }\textbf {\bibinfo {volume} {60}},\ \bibinfo {pages} {309}
  (\bibinfo {year} {1938})}\BibitemShut {NoStop}%
\bibitem [{\citenamefont {Lange}\ \emph {et~al.}(2009)\citenamefont {Lange},
  \citenamefont {Posner}, \citenamefont {Pohl}, \citenamefont {Thierfelder},
  \citenamefont {Grundmeier}, \citenamefont {Blankenburg},\ and\ \citenamefont
  {Schmidt}}]{Lange2009}%
  \BibitemOpen
  \bibfield  {author} {\bibinfo {author} {\bibfnamefont {B.}~\bibnamefont
  {Lange}}, \bibinfo {author} {\bibfnamefont {R.}~\bibnamefont {Posner}},
  \bibinfo {author} {\bibfnamefont {K.}~\bibnamefont {Pohl}}, \bibinfo {author}
  {\bibfnamefont {C.}~\bibnamefont {Thierfelder}}, \bibinfo {author}
  {\bibfnamefont {G.}~\bibnamefont {Grundmeier}}, \bibinfo {author}
  {\bibfnamefont {S.}~\bibnamefont {Blankenburg}},\ and\ \bibinfo {author}
  {\bibfnamefont {W.~G.}\ \bibnamefont {Schmidt}},\ }\bibfield  {title}
  {\bibinfo {title} {{Water adsorption on hydrogenated Si(1 1 1) surfaces}},\
  }\href {https://doi.org/10.1016/j.susc.2008.10.030} {\bibfield  {journal}
  {\bibinfo  {journal} {Surface Science}\ }\textbf {\bibinfo {volume} {603}},\
  \bibinfo {pages} {60} (\bibinfo {year} {2009})}\BibitemShut {NoStop}%
\bibitem [{\citenamefont {Silvestrelli}\ \emph {et~al.}(2006)\citenamefont
  {Silvestrelli}, \citenamefont {Toigo},\ and\ \citenamefont
  {Ancilotto}}]{Silvestrelli2006}%
  \BibitemOpen
  \bibfield  {author} {\bibinfo {author} {\bibfnamefont {P.~L.}\ \bibnamefont
  {Silvestrelli}}, \bibinfo {author} {\bibfnamefont {F.}~\bibnamefont
  {Toigo}},\ and\ \bibinfo {author} {\bibfnamefont {F.}~\bibnamefont
  {Ancilotto}},\ }\bibfield  {title} {\bibinfo {title} {{Interfacial water on
  Cl- and H-terminated Si(111) surfaces from first-principles calculations}},\
  }\href {https://doi.org/10.1021/jp0618781} {\bibfield  {journal} {\bibinfo
  {journal} {Journal of Physical Chemistry B}\ }\textbf {\bibinfo {volume}
  {110}},\ \bibinfo {pages} {12022} (\bibinfo {year} {2006})}\BibitemShut
  {NoStop}%
\bibitem [{\citenamefont {Chen}\ \emph {et~al.}(2018)\citenamefont {Chen},
  \citenamefont {He}, \citenamefont {Liu}, \citenamefont {Qian},\ and\
  \citenamefont {Kim}}]{Chen2018}%
  \BibitemOpen
  \bibfield  {author} {\bibinfo {author} {\bibfnamefont {L.}~\bibnamefont
  {Chen}}, \bibinfo {author} {\bibfnamefont {X.}~\bibnamefont {He}}, \bibinfo
  {author} {\bibfnamefont {H.}~\bibnamefont {Liu}}, \bibinfo {author}
  {\bibfnamefont {L.}~\bibnamefont {Qian}},\ and\ \bibinfo {author}
  {\bibfnamefont {S.~H.}\ \bibnamefont {Kim}},\ }\bibfield  {title} {\bibinfo
  {title} {{Water Adsorption on Hydrophilic and Hydrophobic Surfaces of
  Silicon}},\ }\href {https://doi.org/10.1021/acs.jpcc.8b01821} {\bibfield
  {journal} {\bibinfo  {journal} {Journal of Physical Chemistry C}\ }\textbf
  {\bibinfo {volume} {122}},\ \bibinfo {pages} {11385} (\bibinfo {year}
  {2018})}\BibitemShut {NoStop}%
\bibitem [{\citenamefont {Cao}\ \emph {et~al.}(2011)\citenamefont {Cao},
  \citenamefont {Xu}, \citenamefont {Varghese},\ and\ \citenamefont
  {Heath}}]{Cao2011}%
  \BibitemOpen
  \bibfield  {author} {\bibinfo {author} {\bibfnamefont {P.}~\bibnamefont
  {Cao}}, \bibinfo {author} {\bibfnamefont {K.}~\bibnamefont {Xu}}, \bibinfo
  {author} {\bibfnamefont {J.~O.}\ \bibnamefont {Varghese}},\ and\ \bibinfo
  {author} {\bibfnamefont {J.~R.}\ \bibnamefont {Heath}},\ }\bibfield  {title}
  {\bibinfo {title} {{The microscopic structure of adsorbed water on
  hydrophobic surfaces under ambient conditions}},\ }\href
  {https://doi.org/10.1021/nl2036639} {\bibfield  {journal} {\bibinfo
  {journal} {Nano Letters}\ }\textbf {\bibinfo {volume} {11}},\ \bibinfo
  {pages} {5581} (\bibinfo {year} {2011})}\BibitemShut {NoStop}%
\bibitem [{\citenamefont {Atkins}\ and\ \citenamefont
  {de~Paula}(2014)}]{Atkins2014}%
  \BibitemOpen
  \bibfield  {author} {\bibinfo {author} {\bibfnamefont {P.}~\bibnamefont
  {Atkins}}\ and\ \bibinfo {author} {\bibfnamefont {J.}~\bibnamefont
  {de~Paula}},\ }\href@noop {} {\emph {\bibinfo {title} {{Atkins' Physical
  chemistry}}}},\ \bibinfo {edition} {10th}\ ed.\ (\bibinfo  {publisher}
  {Oxford University Press},\ \bibinfo {year} {2014})\BibitemShut {NoStop}%
\bibitem [{\citenamefont {Barrera}\ \emph {et~al.}(1978)\citenamefont
  {Barrera}, \citenamefont {Guzm{\'{a}}n},\ and\ \citenamefont
  {Balaguer}}]{Barrera1978}%
  \BibitemOpen
  \bibfield  {author} {\bibinfo {author} {\bibfnamefont {R.~G.}\ \bibnamefont
  {Barrera}}, \bibinfo {author} {\bibfnamefont {O.}~\bibnamefont
  {Guzm{\'{a}}n}},\ and\ \bibinfo {author} {\bibfnamefont {B.}~\bibnamefont
  {Balaguer}},\ }\bibfield  {title} {\bibinfo {title} {{Point charge in a
  three-dielectric medium with planar interfaces}},\ }\href
  {https://doi.org/10.1119/1.11501} {\bibfield  {journal} {\bibinfo  {journal}
  {American Journal of Physics}\ }\textbf {\bibinfo {volume} {46}},\ \bibinfo
  {pages} {1172} (\bibinfo {year} {1978})}\BibitemShut {NoStop}%
\bibitem [{\citenamefont {Pont}\ and\ \citenamefont {Serra}(2015)}]{Pont2015}%
  \BibitemOpen
  \bibfield  {author} {\bibinfo {author} {\bibfnamefont {F.~M.}\ \bibnamefont
  {Pont}}\ and\ \bibinfo {author} {\bibfnamefont {P.}~\bibnamefont {Serra}},\
  }\bibfield  {title} {\bibinfo {title} {{ Comment on “Point charge in a
  three-dielectric medium with planar interfaces” [Am. J. Phys. 46 ,
  1172–1179 (1978)] }},\ }\href {https://doi.org/10.1119/1.4907259}
  {\bibfield  {journal} {\bibinfo  {journal} {American Journal of Physics}\
  }\textbf {\bibinfo {volume} {83}},\ \bibinfo {pages} {475} (\bibinfo {year}
  {2015})}\BibitemShut {NoStop}%
\bibitem [{\citenamefont {Aslam}\ \emph {et~al.}(2013)\citenamefont {Aslam},
  \citenamefont {Waldherr}, \citenamefont {Neumann}, \citenamefont {Jelezko},\
  and\ \citenamefont {Wrachtrup}}]{Aslam2013b}%
  \BibitemOpen
  \bibfield  {author} {\bibinfo {author} {\bibfnamefont {N.}~\bibnamefont
  {Aslam}}, \bibinfo {author} {\bibfnamefont {G.}~\bibnamefont {Waldherr}},
  \bibinfo {author} {\bibfnamefont {P.}~\bibnamefont {Neumann}}, \bibinfo
  {author} {\bibfnamefont {F.}~\bibnamefont {Jelezko}},\ and\ \bibinfo {author}
  {\bibfnamefont {J.}~\bibnamefont {Wrachtrup}},\ }\bibfield  {title} {\bibinfo
  {title} {{Photo-induced ionization dynamics of the nitrogen vacancy defect in
  diamond investigated by single-shot charge state detection}},\ }\bibfield
  {journal} {\bibinfo  {journal} {New Journal of Physics}\ }\textbf {\bibinfo
  {volume} {15}},\ \href {https://doi.org/10.1088/1367-2630/15/1/013064}
  {10.1088/1367-2630/15/1/013064} (\bibinfo {year} {2013}),\ \Eprint
  {https://arxiv.org/abs/1209.0268} {arXiv:1209.0268} \BibitemShut {NoStop}%
\bibitem [{\citenamefont {Chandran}\ \emph {et~al.}(2016)\citenamefont
  {Chandran}, \citenamefont {Michaelson}, \citenamefont {Saguy},\ and\
  \citenamefont {Hoffman}}]{Chandran2016}%
  \BibitemOpen
  \bibfield  {author} {\bibinfo {author} {\bibfnamefont {M.}~\bibnamefont
  {Chandran}}, \bibinfo {author} {\bibfnamefont {S.}~\bibnamefont
  {Michaelson}}, \bibinfo {author} {\bibfnamefont {C.}~\bibnamefont {Saguy}},\
  and\ \bibinfo {author} {\bibfnamefont {A.}~\bibnamefont {Hoffman}},\
  }\bibfield  {title} {\bibinfo {title} {{Fabrication of a nanometer thick
  nitrogen delta doped layer at the sub-surface region of (100) diamond}},\
  }\bibfield  {journal} {\bibinfo  {journal} {Applied Physics Letters}\
  }\textbf {\bibinfo {volume} {109}},\ \href
  {https://doi.org/10.1063/1.4971312} {10.1063/1.4971312} (\bibinfo {year}
  {2016})\BibitemShut {NoStop}%
\bibitem [{\citenamefont {Salvadori}\ \emph {et~al.}(2010)\citenamefont
  {Salvadori}, \citenamefont {Ara{\'{u}}jo}, \citenamefont {Teixeira},
  \citenamefont {Cattani}, \citenamefont {Pasquarelli}, \citenamefont {Oks},\
  and\ \citenamefont {Brown}}]{Salvadori2010}%
  \BibitemOpen
  \bibfield  {author} {\bibinfo {author} {\bibfnamefont {M.~C.}\ \bibnamefont
  {Salvadori}}, \bibinfo {author} {\bibfnamefont {W.~W.}\ \bibnamefont
  {Ara{\'{u}}jo}}, \bibinfo {author} {\bibfnamefont {F.~S.}\ \bibnamefont
  {Teixeira}}, \bibinfo {author} {\bibfnamefont {M.}~\bibnamefont {Cattani}},
  \bibinfo {author} {\bibfnamefont {A.}~\bibnamefont {Pasquarelli}}, \bibinfo
  {author} {\bibfnamefont {E.~M.}\ \bibnamefont {Oks}},\ and\ \bibinfo {author}
  {\bibfnamefont {I.~G.}\ \bibnamefont {Brown}},\ }\bibfield  {title} {\bibinfo
  {title} {{Termination of diamond surfaces with hydrogen, oxygen and fluorine
  using a small, simple plasma gun}},\ }\href
  {https://doi.org/10.1016/j.diamond.2010.01.002} {\bibfield  {journal}
  {\bibinfo  {journal} {Diamond and Related Materials}\ }\textbf {\bibinfo
  {volume} {19}},\ \bibinfo {pages} {324} (\bibinfo {year} {2010})}\BibitemShut
  {NoStop}%
\bibitem [{\citenamefont {Iakoubovskii}\ and\ \citenamefont
  {Adriaenssens}(2000)}]{Iakoubovskii2000}%
  \BibitemOpen
  \bibfield  {author} {\bibinfo {author} {\bibfnamefont {K.}~\bibnamefont
  {Iakoubovskii}}\ and\ \bibinfo {author} {\bibfnamefont {G.~J.}\ \bibnamefont
  {Adriaenssens}},\ }\bibfield  {title} {\bibinfo {title} {{Optical transitions
  at the substitutional nitrogen centre in diamond}},\ }\bibfield  {journal}
  {\bibinfo  {journal} {Journal of Physics Condensed Matter}\ }\textbf
  {\bibinfo {volume} {12}},\ \href {https://doi.org/10.1088/0953-8984/12/6/102}
  {10.1088/0953-8984/12/6/102} (\bibinfo {year} {2000})\BibitemShut {NoStop}%
\end{thebibliography}%

\end{document}